\documentclass[aps, prd, twocolumn, superscriptaddress, nofootinbib]{revtex4}
\usepackage{graphicx}
\usepackage{dcolumn}
\usepackage{amssymb}

\begin{document}

\newcommand{\fd}{\ensuremath{f_{10}}}
\newcommand{\ns}{\ensuremath{n_{\mathrm{s}}}}
\newcommand{\Ob}{\ensuremath{\Omega_{\mathrm b}}}
\newcommand{\Oc}{\ensuremath{\Omega_{\mathrm c}}}
\newcommand{\Om}{\ensuremath{\Omega_{\mathrm m}}}
\newcommand{\Obhh}{\ensuremath{\Ob h^{2}}}
\newcommand{\Omhh}{\ensuremath{\Om h^{2}}}
\newcommand{\Ochh}{\ensuremath{\Oc h^{2}}}
\newcommand{\Ol}{\ensuremath{\Omega _{\Lambda}}}
\newcommand{\optdepth}{\eta}
\newcommand{\As}{A_\mathrm{s}}
\newcommand{\Asz}{A_\mathrm{SZ}}
\newcommand{\Cl}{\mathcal{C}_{\ell}}
\newcommand{\Dl}{\mathcal{D}_{\ell}}
\newcommand{\VEV}{\PHI_{0}}
\newcommand{\PHI}{\phi}
\newcommand{\conj}{^*}
\newcommand{\FT}[1]{\tilde{#1}}
\newcommand{\vect}[1]{\mathbf{#1}}
\newcommand{\diff}{\mathrm{d}}
\newcommand{\Sr}{^{\mathrm{S}}}
\newcommand{\Vr}[1]{\!\!\stackrel{\scriptstyle{\mathrm{V}}}{_{\!\!#1}}}
\newcommand{\Tr}[1]{\!\!\stackrel{\scriptstyle{\mathrm{T}}}{_{\!#1}}}
\newcommand{\tauOff}{\tau_{\xi=0}}

\newcommand{\unit}[1]{\;\mathrm{#1}}
\newcommand{\Eq}[1]{Eq.~(\ref{eqn:#1})}
\newcommand{\Eqnb}[1]{Eq.~\ref{eqn:#1}}
\newcommand{\Fig}[1]{Fig.~\ref{fig:#1}}
\newcommand{\Sec}[1]{Sec.~\ref{sec:#1}}
\newcommand{\Table}[1]{Table~\ref{tab:#1}}


\title{CMB power spectra from cosmic strings: predictions\\ 
for the Planck satellite and beyond}

\newcommand{\addressSussex}{Department of Physics \& Astronomy, University of Sussex, Brighton, BN1 9QH, United Kingdom}

\author{Neil Bevis} 
\email{n.bevis@imperial.ac.uk}
\affiliation{Theoretical Physics, Blackett Laboratory, Imperial College, London, SW7 2BZ, United Kingdom}

\author{Mark Hindmarsh} 
\email{m.b.hindmarsh@sussex.ac.uk}
\affiliation{\addressSussex}

\author{Martin Kunz}
\email{martin.kunz@unige.ch}
\affiliation{\addressSussex}
\affiliation{D\'epartement de Physique Th\'eorique, Universit\'e de Gen\`eve, 1211 Gen\`eve 4, Switzerland}
\affiliation{Institut d'Astrophysique Spatiale, Universit\'e Paris-Sud XI, Orsay 91405, France}

\author{Jon Urrestilla}
\email{jon.urrestilla@ehu.es}
\affiliation{Department of Theoretical Physics, University of the Basque Country UPV-EHU, 48040 Bilbao, Spain}
\affiliation{\addressSussex}

\date{May 2010}
\begin{abstract}
We present a significant improvement over our previous calculations of the cosmic string contribution to cosmic microwave background (CMB) power spectra, with particular focus on sub-WMAP angular scales. These smaller scales are relevant for the now-operational Planck satellite and additional sub-orbital CMB projects that have even finer resolutions. We employ larger Abelian Higgs string simulations than before and we additionally model and extrapolate the statistical measures from our simulations to smaller length scales. We then use an efficient means of including the extrapolations into our Einstein-Boltzmann calculations in order to yield accurate results over the multipole range $2\leq\ell\leq4000$. Our results suggest that power-law behaviour cuts in for $\ell\gtrsim3000$ in the case of the temperature power spectrum, which then allows cautious extrapolation to even smaller scales. We find that a string contribution to the temperature power spectrum making up $10\%$ of power at $\ell=10$ would be larger than the Silk-damped primary adiabatic contribution for $\ell\gtrsim3500$. Astrophysical contributions such as the Sunyaev-Zeldovich effect also become important at these scales and will reduce the sensitivity to strings, but these are potentially distinguishable by their frequency-dependence.
\end{abstract}


\maketitle


\section{Introduction}
\label{sec:intro}

Observations of the cosmic microwave background (CMB) radiation and the large-scale distribution of galaxies indicate that cosmic structure was seeded in the very early stages of the universe \cite{Nolta:2008ih,Percival:2009xn}, consistent with the inflationary paradigm. However, the datasets leave room for significant effects due to the presence of cosmic strings at subsequent times \cite{Bevis:2007gh, Battye:2006pk, Fraisse:2006xc, Wyman:2005tu}. These strings (see Refs. \cite{Vilenkin:1994book, Hindmarsh:1994re, Sakellariadou:2006qs, Copeland:2009ga} for reviews) are particularly important in that they are predicted by many physically motivated inflation models, including brane inflation models in the context of string theory \cite{Sarangi:2002yt, Jones:2003da} but also models rooted in Grand Unified Theory (GUT) \cite{Jeannerot:2003qv}. 
Such models also predict other types of cosmic defect, including textures and semilocal strings
\cite{Vachaspati:1991dz,Hindmarsh:1991jq,Achucarro:1999it,Achucarro:2007sp}, and their CMB signals are \cite{Pen:1997ae,Durrer:1998rw,Bevis:2004wk,Cruz:2007pe,Urrestilla:2007sf} also of great interest.

In previous works we calculated cosmic string CMB temperature and polarization power spectra using field theory simulations of the Abelian Higgs model \cite{Bevis:2006mj, Bevis:2007qz} and used the results to fit models with both string- and inflation-induced anisotropies to CMB data \cite{Bevis:2007gh}. It was found that the data (which was primarily the WMAP third-year release \cite{WMAP3-T=Hinshaw:2006ia}) favoured a model with a fractional string contribution to the temperature power spectrum at multipole moment $\ell = 10$ of $\fd = 0.09 \pm 0.05$ with the spectral tilt of the inflation-induced primordial perturbations being $\ns = 1.00 \pm 0.03$. The latter was in contrast to the inflation-only result of $\ns = 0.951^{+0.015}_{-0.019}$ \cite{WMAP3-C=Spergel:2006hy} and showed, along with Refs. \cite{Battye:2006pk,Urrestilla:2007sf}, that the inclusion of topological defects could readily allow $\ns>1$ under those data, thanks to a parameter degeneracy found in Ref. \cite{Bevis:2004wk}.

In the present article we present a significant improvement in the CMB power spectrum predictions from string simulations, with particular emphasis on small angular scales. While the numerical simulations of the Abelian Higgs model in Ref. \cite{Bevis:2006mj} yielded CMB results covering multipoles $\ell \approx 2 \rightarrow 1000$, the full range of angular scales relevant for WMAP data, they did not have the dynamic range to accurately investigate finer angular scales. Higher multipoles are now of great interest thanks to good coverage by sub-orbital CMB experiments 
\cite{CBI=Sievers:2009ah, ACBAR=Reichardt:2008ay, QUAD=Friedman:2009dt, SPT=Lueker:2009rx,ACT=Fowler:2010cy}, plus the imminent arrival of full-sky Planck data \cite{Planck} ($\ell \lesssim 2500$), and we have therefore focused on yielding accurate string predictions for $\ell\lesssim4000$.

Our ability to now include smaller scales stems partially from improvements in computational facilities and the carefully-chosen initial conditions that we now employ in our Abelian Higgs simulations. But more importantly, we have made developments in the method used to yield the CMB power spectra from the statistical measures of the energy-momentum distribution in our simulations: the unequal-time correlation functions (UETCs) \cite{Hindmarsh:1993pu, Pen:1997ae}. We now model and extrapolate the measured UETCs to smaller scales in order to mimic results from larger simulations and have developed an efficient means of including UETC results at these extrapolated scales into our CMB calculations.

The need for the above extrapolation can be understood as follows. The width of a cosmic string is microscopic (perhaps at the GUT scale) while their separation at times of cosmological importance is of order the Hubble distance and hence it is not possible to simultaneously resolve both scales in numerical simulations during the epochs when strings would have impacted on the CMB. We can solve this problem by using \emph{scaling} \cite{Kibble:1984hp}: when the horizon is greater than about one hundred times the string width, we observe that the strings enter a late-time attractor solution in which statistical measures of the string distribution, when measured in horizon units, are constant in time. For example the average string length in a horizon volume divided by the horizon size is $\approx50$. Similarly, the UETC measurements are independent of time once scaled by the horizon size and hence scaling provides knowledge of their values at times of cosmological importance. However, the measured UETC power is attenuated on scales close to the string width and scaling is broken for such length scales. 
Hence the resultant CMB results would show too little power at very small angular scales: the sharp changes in temperature caused by the Gott-Kaiser-Stebbins (GKS) effect \cite{Gott:1984ef, Kaiser:1984iv,Stebbins:1987va} would have been smeared out by having been effectively sourced by strings whose width was around one-hundredth of the horizon size in the post-recombination era. For our previous work this was of little actual concern since those articles focused upon WMAP scales ($\ell\lesssim1000$). It is, however, necessary to extrapolate the UETC results to sub-string scales in order to obtain accurate results from field theoretic simulations for much higher multipoles. 

Other approaches to calculating CMB perturbations from strings include using simulations of the Nambu-Goto (NG) type \cite{Contaldi:1998mx, Landriau:2003xf, Fraisse:2007nu} and the unconnected segment model (USM) \cite{Albrecht:1997nt, Pogosian:1999np, Battye:2006pk, Pogosian:2008am}. Nambu-Goto simulations can yield a greater dynamic range than field theoretic simulations but must still invoke scaling to yield CMB results over a wide range of scales. Furthermore the network correlation length used in the initial conditions appears to persist and may limit the reliable resolution of Nambu-Goto simulations \cite{Ringeval:2005kr, Olum:2006ix}. The USM represents the string network as a stochastic set of moving ``sticks", which disappear at an appropriate rate in order to give the chosen string scaling density. While it has no true dynamical content, it is computationally cheap and offers the flexibility to choose the coarse-grained network properties \cite{Pogosian:1999np,Battye:2006pk,Pogosian:2007gi,Battye:2010hg,Battye:2010xz}.

The advantage of the Abelian Higgs model is that it includes the small-scale physics near the string width, which has a non-negligible impact on the string dynamics \cite{Vincent:1997cx, Hindmarsh:2008dw}: energy from the strings is converted into massive gauge and Higgs radiation. In NG simulations this decay channel is not included and the string length density is significantly higher, with the long strings being converted into small loops that would then decay via gravitational radiation (although this process is not actually simulated). With the extra decay channel in the field theoretic simulations, decay via gravitational radiation would be less important, and this fact significantly changes the predictions for gravitational wave observations. In the case of the CMB on the other hand, it is the long strings that are important, and the key difference between the simulation results is the inter-string separation. A potential disadvantage of a field theory simulation is that computational constraints require the string width to be artificially increased in order to keep it above the simulation resolution, but we carefully show that this does not significantly affect the UETCs and therefore the CMB power spectra results.

In Sections \ref{sec:method} and \ref{sec:refine} we detail our methods, including an overview of the UETC approach, our field theory simulations, the tests of scaling that we employ and the sub-string extrapolation.
We exhibit the resulting CMB power spectra in Section \ref{sec:CMBresults}, and give conclusions in Section \ref{sec:conclusions}. 


\section{Method overview}
\label{sec:method}

As already noted, the basis for our string simulations and CMB calculations is our previous work: Ref. \cite{Bevis:2006mj}, referred to hereafter as BHKU. We refer the reader to that article for the full details of our approach but we present the essential information in what immediately follows and highlight the improvements made for the current article in the next section.


\subsection{UETC approach}

In CMB power spectra calculations for inflationary models with cosmic strings, the inflationary contribution is essentially uncorrelated with the cosmic string contribution. This is because the inflationary perturbations are laid down by an independent field $60$ e-foldings before the strings are formed, and because the complex string dynamics rapidly destroys correlations with earlier times (see \Sec{UETCresults}). As a result we may write the total spectrum $\ell(\ell+1) \Cl$ as a sum of two independent spectra and take the cross-correlation term to be negligible:
\begin{equation}
\label{eqn:powerSum}
\Cl = \Cl^{\mathrm{inf}} + \Cl^{\mathrm{str}}, 
\end{equation}
where $\Cl^{\mathrm{inf}}$ is the inflationary contribution and $\Cl^{\mathrm{str}}$ is the string contribution. We can use standard methods \cite{Seljak:1996is} to determine $\Cl^{\mathrm{inf}}$ and it is therefore upon $\Cl^{\mathrm{str}}$ that this article is focused. 

Physically the strings cause CMB anisotropies by creating perturbations in the space-time metric, which are roughly of the same order as $G\mu$, where $\mu$ is the string mass per unit length, $G$ is the gravitational constant, and $G\mu\lesssim10^{-6}$ to be consistent with current observations \cite{Bevis:2007gh}. These inhomogeneities in the metric then lead to perturbations in the matter and radiation that themselves evolve and influence the strings, but we can neglect this back-reaction since the resulting perturbations of the strings would then result in changes only of order $(G\mu)^2$ to the metric.

In order to determine the string contribution to a two-point correlation function, such as the CMB temperature power spectrum, we are required to solve a set of linear differential equations of the following form, in which the string energy-momentum tensor components act as source terms $\FT{S}_{a}$:
\begin{equation}
 \hat{\mathcal{D}}_{ac}(k,a,\rho,...)
 \FT{X}_{a}(\vect{k},\tau)
 \; = \;
 \FT{S}_{c}(\vect{k},\tau).
\end{equation}
Here $X_a$ is the quantity of interest and $\FT{X}_{a}$ its Fourier transform, while $\hat{\mathcal{D}}_{ac}$ is a differential operator, dependent upon the cosmic scale factor $a$, the background matter density $\rho$ and similar quantities. Our notation is such that $\tau$ is the conformal time and $\vect{k}$ is the comoving wavevector. The homogeneous version of this equation $(\FT{S}_{c}=0)$, which corresponds to the inflationary case, can be solved by standard codes and therefore, in principle, we may use a Green's function $\mathcal{G}_{ac}(k,\tau_0,\tau)$ to give the power spectrum at conformal time $\tau_0$ for the string case via:
\begin{eqnarray}
 \label{eqn:Greens}
 \left<\! 
   \FT{X}_{a}(\vect{k},\tau_0) \FT{X}_{b}\conj(\vect{k},\tau_0)
 \!\right> \! = \! 
 \!\int_{0}^{\tau_0}\!\!\!\! \int_{0}^{\tau_0} \!\!\!\!\!\!
 \diff\tau \diff\tau' & & \!\!\!\!\!\! 
 \mathcal{G}_{ac}(k,\tau_0,\tau\!) \mathcal{G}_{bd}\conj(k,\tau_0,\tau'\!)
 \nonumber \\ && \!\!\!\!\!\!\!\!
    \times \left<\!
      \FT{S}_{c}(\vect{k},\tau\!) \FT{S}_{d}\conj(\vect{k},\tau'\!)
    \!\right>\!.
\end{eqnarray}
Hence the data required to calculate such two-point correlation functions are the two-point unequal-time correlators of the string energy-momentum tensor components \cite{Hindmarsh:1993pu, Pen:1997ae}:
\begin{equation}
 \FT{U}_{ab}(k,\tau,\tau')
 = 
 \left< \FT{S}_{a}(\vect{k},\tau) \, \FT{S}_{b}\conj(\vect{k},\tau') \right>.
\end{equation}
Note that statistical isotropy implies that $\FT{U}_{ab}$ is not dependent on the direction of $\vect{k}$ and further that it is real-valued.

Significant simplification in the form of $\FT{U}_{ab}$ may be made using the scaling property, briefly mentioned in the introduction. Under scaling, any statistical measure of the spatial distribution of strings scales with the horizon size, which is just $\tau$ in comoving coordinates. For example the comoving length-density of string is $\alpha/\tau^2$, where $\alpha$ is a dimensionless constant. The existence of this attractor was predicted by Kibble \cite{Kibble:1984hp} and has been confirmed in Nambu-Goto simulations \cite{Fraisse:2007nu, Olum:2006ix,Martins:2005es} (with some assumptions about the decay of loops) and in Abelian Higgs simulations \cite{Vincent:1997cx, Moore:2001px, Bevis:2006mj}. We will present further evidence in support of scaling in our results section. 

The power of scaling is that it enables us to write $\FT{U}_{ab}$ in terms of a function of just two variables:
\begin{eqnarray}
 \label{eqn:scalingFunction}
 \FT{U}_{ab}(\vect{k},\tau,\tau')
 & = &
 \frac{\VEV^{4}}{\sqrt{\tau\,\tau'}} \frac{1}{V}
 \;
 \FT{C}_{ab}(k\sqrt{\tau\,\tau'},\tau/\tau').
\end{eqnarray}
Here $\VEV$ sets the energy-scale of the problem and converts between the scaling spatial distribution of string and the distribution of energy. For example, in the case of the Abelian Higgs model (see next section) it is the vacuum expectation value of the scalar field. The comoving simulation volume $V$ appears here because we define our Fourier transform so that it leaves dimensions unchanged:
\begin{equation}
 \FT{X}(\vect{k})
 = 
 \frac{1}{V} 
 \int \diff^{3} \vect{x} \;
  X(\vect{x}) \;
  e^{-i \vect{k}\cdot\vect{x}}.
\end{equation}
The UETC scaling function $\FT{C}_{ab}$ can be seen to allow data to be taken for very large $k$ at small $\tau$ and $\tau'$ and then used to provide information about small $k$ at large $\tau$ and $\tau'$. This is critical for the present calculations, as noted in the introduction. 

Further power of the UETC scaling functions derives from their functional form: they decay for large and small time ratios $\tau/\tau'$, and for large $k\sqrt{\tau\,\tau'}$ (small scales), while causality constrains their form at low $k\sqrt{\tau\,\tau'}$. This crucially means that we need study a scaling network for only a relatively short range of times and in a limited simulation volume. However as noted in the introduction, small length scales become more important when considering small angular scales (see next section).

While our approach is to measure $\FT{C}_{ab}$ from cosmic string simulations, and therefore find $\FT{U}_{ab}$, we do not in fact then use \Eq{Greens}. Instead we decompose $\FT{C}_{ab}$ into a sum over products as \cite{Turok:1996ud}:
\begin{equation}
\FT{C}_{ab} (k\sqrt{\tau\tau'},\tau/\tau') = \sum_{n} \lambda_{n} \FT{c}_{na}(k\tau) \FT{c}_{nb}(k\tau'). 
\end{equation}
The problem then breaks down to:
\begin{equation}
 \label{eqn:Greens2}
 \left<\! 
   \FT{X}_{a}(\vect{k},\tau) \FT{X}_{b}\conj (\vect{k},\tau')
 \!\right>
 =
 \frac{\VEV^{4}}{V} \sum_{n}
 \lambda_{n}
 I^{n}_{a}(k,\tau) \,
 I^{n}_{b}{\conj}(k,\tau'), 
\end{equation}
where:
\begin{equation}
 I^{n}_{a}(k,\tau_0)
 =
 \int^{\tau_0}_0\! \diff\tau \mathcal{G}_{ab}(k,\tau_0,\tau) \, \frac{\FT{c}_{nb}(k\tau)}{\sqrt{\tau}}.
\end{equation}
In practice we do not calculate this integral via a Green's function, but instead apply a modified version of CMBEASY \cite{Doran:2003sy} to determine the CMB power spectrum contribution from the coherent active source $\FT{c}_{nb}/\sqrt{\tau}$.

In principle there are 55 possible UETCs between the 4 scalar, 4 vector and 2 tensor degrees of freedom in the energy-momentum tensor $\FT{T}_{\mu\nu}$, but thanks to statistical isotropy and energy conservation, we in fact need only to measure 5 scaling functions \cite{Durrer:1997ep}. The scalar UETCs that we calculate involve projections from $\FT{T}_{\mu\nu}$ that, via Einstein's equations, directly source the two Bardeen potentials \cite{Bardeen:1980kt}:
\begin{eqnarray}
 \FT{S}\Sr_{\Phi} 
 & = &
 \FT{T}_{00} - 3 \frac{\dot{a}}{a} \frac{i\hat{k}_{m}}{k} \FT{T}_{0m},
\\
 \label{eqn:defPsi}
 \FT{S}\Sr_{\Psi}
 & = &
 -\FT{S}\Sr_{\Phi} - T_{mm} + 3 \hat{k}_{m}\hat{k}_{n} \FT{T}_{mn}.
\end{eqnarray}
From these projections there are 3 independent UETC scaling functions that we must measure:
\newcommand{\removeLHspace}{\!\!\! \!\!\! \!\!\!}
\begin{eqnarray}
 \removeLHspace
 \left<\! 
  \FT{S}\Sr_{\Phi}(\vect{k},\tau) \,
  \FT{S}_{\Phi}^{\mathrm{S}*}(\vect{k},\tau') 
 \!\right>
 & \!\!\!=\!\!\!&
 \frac{\VEV^{4}}{\sqrt{\tau\,\tau'}} \frac{1}{V}
 \FT{C}\Sr_{11}(k\sqrt{\tau\,\tau'},\tau / \tau'), 
\label{eqn:C11}
\\
 \removeLHspace
 \left<\! 
  \FT{S}\Sr_{\Phi}(\vect{k},\tau) \,
  \FT{S}_{\Psi}^{\mathrm{S}*}(\vect{k},\tau') 
 \!\right>
 & \!\!\!=\!\!\! &
 \frac{\VEV^{4}}{\sqrt{\tau\,\tau'}} \frac{1}{V} 
 \FT{C}\Sr_{12}(k\sqrt{\tau\,\tau'},\tau / \tau'), 
\label{eqn:C12}
\\
 \removeLHspace
 \left<\! 
  \FT{S}\Sr_{\Psi}(\vect{k},\tau) \, 
  \FT{S}_{\Psi}^{\mathrm{S}*}(\vect{k},\tau') 
 \right>
 & \!\!\!=\!\!\! &
 \frac{\VEV^{4}}{\sqrt{\tau\,\tau'}} \frac{1}{V} 
 \FT{C}\Sr _{22}(k\sqrt{\tau\,\tau'},\tau / \tau').
\label{eqn:C22}
\end{eqnarray}
Then in the tensor case we project out the two tensor degrees of freedom (see BHKU), which we denote as $\FT{S}^{\mathrm{T1}}$ and $\FT{S}^{\mathrm{T2}}$. We then determine:
\begin{eqnarray}
 \removeLHspace\!\!
 \left<\! 
  \FT{S}^{\mathrm{T1}}\!(\vect{k},\tau) \, 
  \FT{S}^{\mathrm{T1}^*}\!(\vect{k},\tau') 
 \!\right>
 & \!\!\!=\!\!\! &
 \left<\! 
  \FT{S}^{\mathrm{T2}}\!(\vect{k},\tau) \, 
  \FT{S}^{\mathrm{T2}^*}\!(\vect{k},\tau')  \!\right>
\\
 = & & \!\!\!\!\!\! 
 2 \frac{\VEV^{4}}{\sqrt{\tau\,\tau'}} 
 \frac{1}{V}
 \FT{C}\Tr{}\!\!(k\sqrt{\tau\,\tau'},\tau / \tau'),\nonumber 
 \label{eqn:CT}
\end{eqnarray}
where the factor of $2$ is present to ensure that $\FT{C}\Tr{}$ matches the definition of Ref. \cite{Durrer:1998rw}.

Following Ref. \cite{Durrer:1998rw}, we make a change in the scaling function definition when considering the two vector degrees of freedom from $\FT{T}_{0i}$, $\FT{S}^{\mathrm{V1}}$ and $\FT{S}^{\mathrm{V2}}$ (see BHKU). Specifically we pull out a factor of $k^{2}\tau\tau'$ from the scaling function such that:  
\begin{eqnarray}
 \removeLHspace\!\!
 \left<\! 
  \FT{S}^{\mathrm{V1}}\!(\vect{k},\tau) \, 
  \FT{S}^{\mathrm{V1}^*}\!(\vect{k},\tau') 
 \!\right>
 & \!\!\!=\!\!\! &
 \left<\! 
  \FT{S}^{\mathrm{V2}}\!(\vect{k},\tau) \, 
  \FT{S}^{\mathrm{V2}^*}\!(\vect{k},\tau')  \!\right>
  \label{eqn:CV}
  \\
  = & & \!\!\!\!\!\! 
 k^{2}\sqrt{\tau\,\tau'} \VEV^{4} 
 \frac{1}{V}
 \FT{C}\Vr{}\!\!(k\sqrt{\tau\,\tau'},\tau / \tau').\nonumber 
\end{eqnarray}
This definition is motivated by the fact that covariant energy-momentum conservation requires that the vector UETC varies as $k^{2}$ at small $k$ while the other UETCs that we measure tend to a constant value as $k\rightarrow0$; and it is desirable for all UETC scaling functions to have the same super-horizon properties. 

However, it should be noted that the above discussion requires a small change because scaling is broken near the time of radiation-matter equality $\tau_{\mathrm{eq}}$, since $\tau_{\mathrm{eq}}$ is a second dimensional scale which enters the problem. We hence are required to take UETC data in both the radiation and matter eras --- although the matter era data dominates the CMB results --- and we then use interpolation in order to model the transition. Scaling is also broken as the Universe enters the current accelerating phase, causing the string density to decay, which we model by partially suppressing the stress source.

For more details of the solution of the linearized Einstein-Boltzmann equations in the presence of sources see Refs.\ \cite{Hu:1997hp,Durrer:2001cg}.


\subsection{Field theoretic simulations}

For computational speed, we simulate local cosmic strings by solving the classical field equations of the simplest theory that contains them: the Abelian Higgs model. In the notation of BHKU this has Lagrangian density:
\begin{equation}
 \label{eqn:lagrangian}
 \mathcal{L}
 = 
 - \frac{1}{4e^{2}} F_{\mu\nu} F^{\mu\nu}
 + (D_{\mu} \PHI)\conj (D^{\mu} \PHI)
 - \frac{\lambda}{4} \left( |\PHI|^{2} - \VEV^{2} \right)^{2}\!\!,
\end{equation}
with $D_{\mu} = \partial_{\mu} + i A_{\mu}$ and $F_{\mu\nu}=\partial_{\mu} A_{\nu} - \partial_{\nu} A_{\mu}$ while $e$ and $\lambda$ are dimensionless coupling constants. In a spatially flat Friedmann-Robertson-Walker(FRW) metric with scale factor $a$, this leads to the following Euler-Lagrange equations:
\begin{eqnarray}
 \label{eqn:AHdynamicPhi}
 \ddot{\PHI} + 2 \frac{\dot{a}}{a}\dot{\PHI} - D_{j} D_{j} \PHI 
 & = &
 -a^{2} \frac{\lambda}{2} \left( |\PHI|^{2} - \VEV^{2} \right) \PHI,
\\
 \label{eqn:AHdynamicF}
 \dot{F}_{0j} - \partial_{i}F_{ij}
 & = &
 -2a^{2} e^{2} \; \mathcal{I}m \! \left[ \PHI\conj D_{j}\PHI \right],
\\
 \label{eqn:AHconstraint}
 -\partial_{i} F_{0i} & = & -2 a^{2} e^{2} \; \mathcal{I}m [ \PHI\conj \dot\PHI ],
\end{eqnarray}
where the gauge choice $A_{0}=0$ has been made. Here we use over-dots to denote differentiation with respect to conformal time $\tau$, while $\partial_{i}$ denotes differentiation with respect to comoving Cartesian coordinates.

When simulating this model in an expanding universe, the simulations must resolve the comoving string width $w_0/a$ (i.e. $w_0$ is the fixed physical width $\sim\VEV^{-1}$). They must also contain at least one horizon volume of comoving diameter $2\tau$. However, these two scales diverge rapidly, with their ratio growing as $\tau^{3}$ in the matter era, while we require ratios $\gtrsim100$ in order for strings to scale. Hence with $1024^3$ lattice simulations, which are the largest that are practical with our available facilities, we cannot study a scaling network using these equations for long enough to measure the UETCs up to sufficiently high $\tau/\tau'$ ratios. Furthermore, the increase in this ratio with computer time $t_{\mathrm{cpu}}$ varies as $t_{\mathrm{cpu}}\!\!\!\!\!\!\!^{1/12}$ and hence the returns from much larger outlays are minimal. In BHKU we therefore proceeded by allowing temporal variations in the coupling constants $\lambda$ and $e$:
\begin{eqnarray}
\lambda & = & \frac{\lambda_{0}}{a^{2(1-s)}},\\
e       & = & \frac{e_{0}}{a^{1-s}},
\end{eqnarray}
such that the comoving string width now varies as
\begin{equation}
w =  \frac{w_{0}}{a^{s}},\\
\end{equation}
where $s$ is the string width control parameter. If $s=0$, then the factors of $a$ on the right-hand-side of Eqs. (\ref{eqn:AHdynamicPhi})-(\ref{eqn:AHconstraint}) are removed and the string width remains constant in comoving coordinates, while $s=1$ gives the normal dynamics of the model. Note that the above dependencies preserve the ratio $\lambda/2e^2$, which we set to be unity for the simulations described in this article: the Bogomol'nyi limit \cite{Bogomolnyi:1976}.

Simply evolving the above dynamical equations (Eqs. \ref{eqn:AHdynamicPhi} and \ref{eqn:AHdynamicF}) while varying $\lambda$ and $e$ will not preserve the constraint \Eq{AHconstraint}. Hence the BHKU method is to vary the model action with respect to the fields whilst allowing for the temporal dependence in $\lambda$ and $e$, which then yields a consistent set of dynamical and constraint equations: Eqs. (\ref{eqn:AHdynamicPhi}) and (\ref{eqn:AHconstraint}) in addition to a modified form of \Eq{AHdynamicF}:
\begin{equation}
 \label{eqn:AHdynamicFs}
 \dot{F}_{0j} +2(1-s)\frac{\dot{a}}{a}F_{0j} - \partial_{i}F_{ij}
 =
 -2a^{2} e^{2} \; \mathcal{I}m \! \left[ \PHI\conj D_{j}\PHI \right].
\end{equation}
However, for $s\neq1$ the action is no longer a 4-scalar and hence there is a breach of covariant energy conservation. Since it is the energy-momentum tensor that seeds the cosmological perturbations which result from strings, then this is clearly a potential problem. Fortunately in BHKU it was established that the effects on the UETCs, and therefore the CMB power spectra, are minimal on the relevant scales, but we present additional evidence in support of this approach in the next section.


\section{Method refinements and intermediate results}
\label{sec:refine}


As discussed in the introduction, the use of scaling to translate simulation results from GUT length-scales to cosmological scales means that any effects present in our simulations on scales close to the string width are erroneously transfered to scales of order one hundredth of the horizon size at the decoupling of the CMB, i.e. $\sim 3 \unit{kPc}$ rather than $\sim 10^{-32} \unit{m}$ (which would correspond to the string width if $G\mu\sim10^{-6}$). As already mentioned, this smearing out of the energy density would lead to reduced power in the string contribution to the CMB power spectrum at small angular scales. 

For our previous work, which was intended to be used only with WMAP data, this effect was not anticipated to be significant, since WMAP only probes scales larger than about one tenth of the horizon at decoupling ($\ell<1000$). Additionally the small-scale data which then existed was not precise enough or on small enough scales for the likely inaccuracies to be a cause of concern, unless the strings completely dominated the CMB on such scales --- something that the WMAP data was seen to rule out \cite{Bevis:2007gh}. While it is true that at times long after recombination, the properties of strings on scales much smaller than the horizon can influence WMAP scales, their effect would have been minor, as we shall demonstrate in this article, which now includes their contribution.

In order to include small-scale UETC power, we model the UETCs and extrapolate the trends seen to sub-string scales (\Sec{extrapSmallScale}). We have also modified our CMB calculation method in order to rapidly include these extrapolations (\Sec{eigenDecomp}). However, we additionally employ larger simulations than in the past and employ different initial conditions, both of which enhance our ability to study the scaling epoch and improve our measurements of the UETC scaling functions. We hence discuss our new results for measures of scaling and of the UETCs themselves, before going on to discuss the inclusion of small scales.


\subsection{Tests of scaling}
\label{sec:initialConditions}

As has already been alluded to, for strings to scale their width must be much less than their separation, which is of order the horizon size. This statement may be made more precise by introducing the (comoving) \emph{network length scale} $\xi$, defined by:
\begin{equation}
\xi = \sqrt{\frac{V}{L}}, 
\end{equation}
where $L$ is the comoving string length\footnote{We measure $L$ by detecting the lattice grid-squares around which the phase $\PHI$ has a net winding (using the gauge-invariant method of Ref. \cite{Kajantie:1998bg}) and then we approximately reconstruct the string path as a collection of perpendicular segments of length $\Delta x$ passing through these squares.
We then apply the Scherrer-Vilenkin correction factor of $\pi/6$ \cite{Scherrer:1997sq} in order to approximately account for over-estimate due to representing a smooth path by perpendicular segments (see Ref. \cite{Hindmarsh:2008dw} for more discussion).} in the simulation volume $V$. We observe in our simulations that accurate scaling sets in at $\xi\gtrsim40w$ whereupon $\xi\sim0.3\tau$.

\begin{figure}
\resizebox{\columnwidth}{!}{\includegraphics{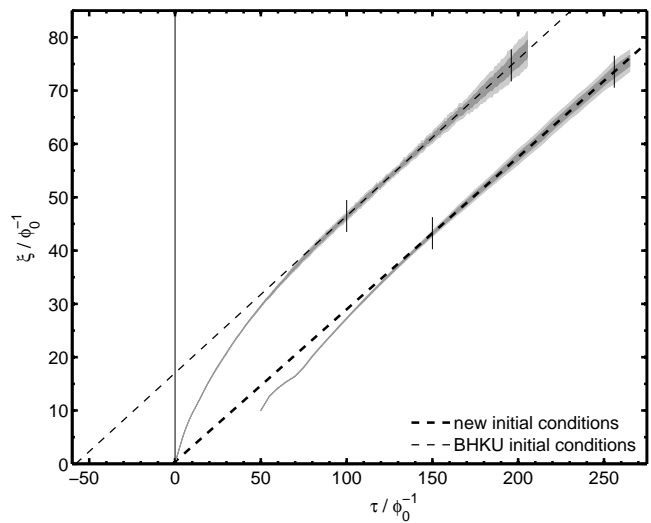}}
\caption{\label{fig:xiCombine}Results for $\xi$ from simulations using our new initial conditions compared to those using BHKU initial conditions, which yielded an offset scaling law. Statistical uncertainties determined from 3 realizations are denoted using the shaded regions ($1-\sigma$ dark, $2-\sigma$ light) and the linear best-fits over the period between the short vertical lines are indicated by the dashed lines. The results come from $1024^{3}$, $s=0$, \mbox{$\Delta x=0.5\VEV^{-1}$} simulations in the case of the new initial conditions and $768^{3}$, $s=0.3$, $\Delta x=0.4295 \VEV^{-1}$ for the BHKU ones.}
\end{figure}

However, in BHKU we were limited to an Abelian Higgs simulation size of $512^{3}$, which is about the minimum required to study a scaling network of strings. We therefore were forced to be content with initial conditions that yielded not $\xi\propto\tau$ but $\xi\propto(\tau-\tauOff)$, as shown in \Fig{xiCombine}, were $\tauOff$ is negative. That is, at early times string decay occurred more quickly than scaling would have predicted and $\xi$ increased rapidly, but then stabilized to an approximately constant $\diff\xi/\diff\tau$ once $\xi\gtrsim40w$ (where $w\approx\VEV^{-1}$ but is slowly decreasing with $\tau$ in the $s=0.3$ simulation for which BHKU-style results are plotted). As can be seen in the figure, this rapid increase aids the creation of an approximately scaling network for simulations that are causally limited to small times, albeit scaling with $(\tau-\tauOff)$ rather than $\tau$. Fortunately, at times of cosmological interest we have $\tau\gg\tauOff$ and therefore this offset is not actually relevant. Furthermore, when $\tau$ was replaced by $(\tau-\tauOff)$ in the calculations of the UETC scaling functions, the equal-time scaling functions $\FT{C}(k\tau,1)$ were seen to be approximately constant in $\tau$, as is required for scaling.

Despite the success of the $(\tau-\tauOff)$ replacement when using the BHKU initial conditions, with the larger $1024^{3}$ simulations available for the present work we noted a slow drift in the $\diff\xi/\diff\tau$ value, even at late times. Further, there was a slow rise in the magnitude of the \mbox{$\tauOff$-corrected} equal-time scaling functions. This is expected at some level since the dynamical equations contain the quantity:
\begin{equation}
\frac{\dot{a}}{a} = \frac{n}{\tau},
\end{equation}
where $n$ is unity for the radiation era and 2 for the matter era. The ratio of this damping scale $\tau/n$ to the network length scale $\xi$ changes with time if $\tauOff\neq0$ and at $\tau\sim|\tauOff|$ with negative $\tauOff$, the system has a lower than asymptotic value of $\tau/n\xi$ --- it is being damped too heavily. We hence now employ initial conditions that yield (effectively) $\tauOff=0$, although now we do not see scaling until long after the causal run-time limit of our previous $512^{3}$ simulations.

In principle, the initial conditions for the fields are set by their fluctuations (quantum or thermal) at string formation, which have a finite, microscopic, correlation length (see e.g.\ \cite{Rajantie:2003xh}). Fortunately the string network is observed to relax to scaling for a wide variety of initial conditions and so it is not important to model the initial fluctuations precisely. In BHKU we employed initial conditions that were designed to model a vacuum phase transition at the end of inflation: each lattice site was given an independent phase for $\PHI$, with $|\PHI|=\VEV$, and the initial time $\tau_\mathrm{start}$ was set roughly equal to the lattice spacing $\Delta x$. The gauge field and the canonical momentum $\dot\PHI$ were set to zero. However, that set of initial conditions means that the simulations begin with $\tau_\mathrm{start}\sim w$ and this was seen in BHKU to be responsible for the initial rise in $\xi$. 
 
In this work we begin with $\PHI$ as a Gaussian random field with correlation length $l_\PHI$ such that \mbox{$l_\PHI\sim\tau_\mathrm{start}\gg w$}. We are free to choose the two-point auto-correlation function of this Gaussian random field, subject to the constraints that it is zero outside the causal horizon and that its Fourier transform (the power spectrum) must be non-negative. For simplicity we relax the first constraint slightly and opt for:
\begin{equation}
P(y) = \frac{1}{V} \!\!\int\!\! \diff x^{3} \PHI(x) \PHI\conj(\vect{x}-\vect{y}) = P_0 \exp(-y^{2}/2l_{\PHI}^{2}).
\end{equation}
This decays so rapidly outside the horizon that it is effectively causal with horizon size $\sim l_{\PHI}$. We hence generate a field in Fourier space with spectrum $\FT{P}$ and random phases for each $\vect{k}$-mode, before transforming it to real space to become our initial $\PHI$ configuration. This gives $\left< \PHI\conj \PHI \right> = P_0$, which we take as $\VEV^{2}$ since the field should be close to its vacuum except at lattice sites very near to the strings. Further, since Hubble damping at $\tau\gg w$ is too weak to rapidly relax the fields into a network of string, we now employ diffusive (first-order) evolution until a time $\tau_\mathrm{diff}$ in order to achieve that. We then have three parameters $\tau_\mathrm{start}$, $\tau_\mathrm{diff}$ and $l_\PHI$ which we may vary in order to achieve our goal of $\tauOff=0$, although each test of a point in this 3-dimensional parameter space is very computationally-expensive.

While in principle this $\tauOff\rightarrow0$ optimization should be performed for each value of $s$, the insensitivity of the string dynamics to changes in $s$ (as noted in BHKU) implies that this computationally-costly process can be performed only once. However, to keep the same settings for these 3 parameters at all $s$ values, we must use a lower lattice spacing $\Delta x$ at higher $s$ in order for strings to still be resolved ($w_0/a^s>2\Delta x$) at the end of the simulation $\tau=N\Delta x/2$, which is the causal time limit\footnote{We continue the simulations slightly beyond the strict causal limit, since they do not feel the periodic boundary conditions immediately, but the further reduction in string width is not sufficient to greatly affect the reliability of the simulations.}. This implies that the accessible $\tau/\tau'$ range narrows as $s$ is increased.

\begin{table}
\begin{ruledtabular}
\begin{tabular}{cccc}
Simulation & Measure          & Radiation era   & Matter era \\
\hline
AH & $\xi/\tau$         & $0.255\pm0.018$ & $0.285 \pm 0.011$ \\
AH & $L\tau^2/V$        & $15\pm2$        & $12.2 \pm 0.96$ \\ 
NG & $L\tau^2/V$ & $37.8 \pm 1.7$ & $28.4\pm0.9$ \\
\end{tabular}
\end{ruledtabular}
\caption{\label{tab:stringXi} 
Numerical results for the network length scale $\xi$ in horizon units and the string length density $L/V$ normalized to the horizon size. AH indicates that the results are from the present Abelian Higgs simulations while NG indicates the results are from the Nambu-Goto simulation of Ref. \cite{Ringeval:2005kr}(see \Sec{compNG}), which yield string densities approximately $2.5$ (radiation) and $2.3$ (matter) times greater. (Note the NG results quoted includes only infinite strings, although loops do not contribute significantly to our figures \cite{Hindmarsh:2008dw}.)}
\end{table}
\begin{figure}
\resizebox{\columnwidth}{!}{\includegraphics{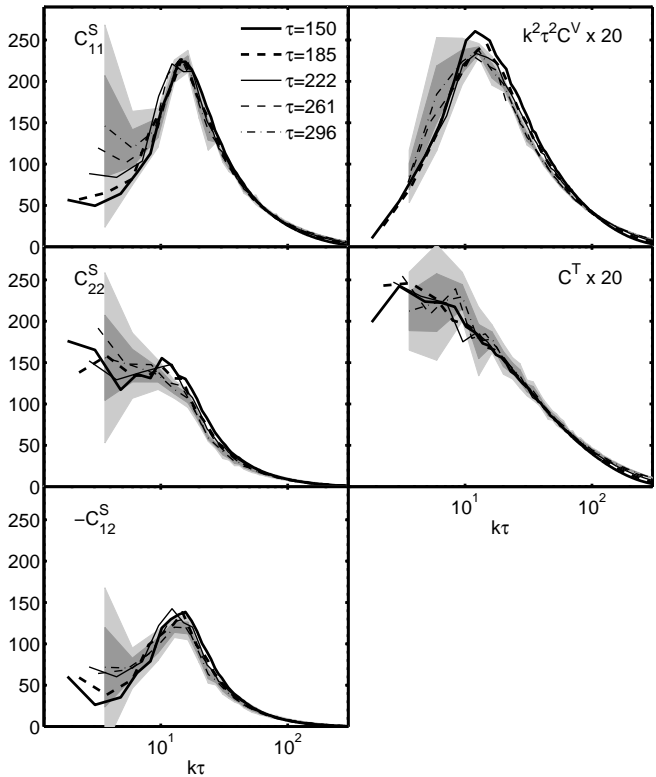}}
\caption{\label{fig:ETCscaling}The equal-time scaling functions $\FT{C}(k\tau,1)$ from $1024^{3}$, $s=0$ simulations with a matter-dominated FRW background for 5 times, as indicated in the legend in units of $\VEV^{-1}$. Results shown are the average of 3 realizations, with the estimated statistical uncertainties for $\tau=296\VEV^{-1}$ indicated by the shaded regions.}
\end{figure}
The results for $\xi$ from these new initial conditions can be seen in \Fig{xiCombine}, for simulations in the matter era using $s=0$ (with $a=1$, $\lambda=2$ and $e=1$ during diffusive evolution and $\Delta x=0.5\VEV^{-1}$). Further, we show from the same simulations examples of the equal-time scaling functions $\FT{C}(k\tau,1)$ in \Fig{ETCscaling}, which provide a scale-dependent test of scaling and is the quantity that is most important for the CMB calculations. These results exhibit scaling on large and intermediate scales to a good degree of accuracy over a ratio in conformal time of 2. A breach of scaling is evident at small-scales, where statistical uncertainties are highly suppressed, but as noted in the introduction this is expected due to the proximity of the associated length scales to the string width. We discuss this issue and our solution to it in \Sec{extrapSmallScale}. Numerical results for the string length density under scaling are also given in \Table{stringXi} (see also \Sec{compNG}). 


\subsection{Dependence of the equal-time scaling functions upon $s$}
\label{sec:ETCvaryS}

\begin{figure}
\resizebox{\columnwidth}{!}{\includegraphics{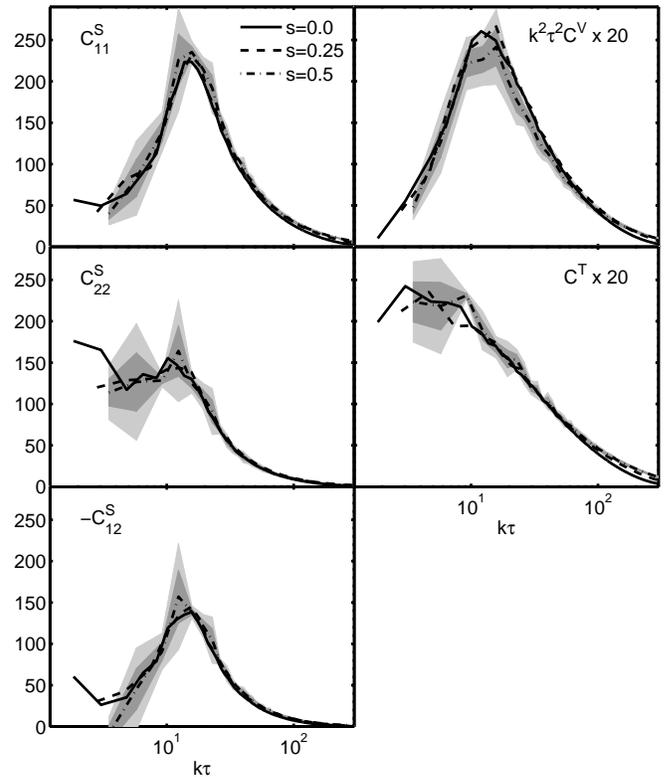}}
\caption{\label{fig:ETCvaryS}The equal-time scaling functions $\FT{C}(k\tau,1)$ from simulations with a matter-dominated FRW background at three values of $s$. Results shown are the average of 3 realizations, with statistical uncertainties for the $s=0.5$ case indicated by the shaded regions. Results are plotted for $\tau=150\VEV^{-1}$, which is the start of the period when we take UETC data for the $s=0$ case.}
\end{figure}

In addition to ensuring that the simulations are exhibiting scaling, we must also check that the use of $s<1$ in our dynamical equations does not introduce significant systematic errors in our UETC scaling function results. This is most readily illustrated using the equal-time case again, with results shown in \Fig{ETCvaryS}. We can resolve no dependence upon $s$, except for at high $k\tau$, which is again due to the proximity of the corresponding length scales to the string width. Since the affected scales are the very ones for which we apply the extrapolation mentioned in the introduction, then this effect is of no concern (see \Sec{extrapSmallScale}). For completion, however, the $s$-dependence on these scales can be explained as follows. The higher $s$, the more rapid the reduction in comoving string width during the Hubble phase and therefore the attenuation of small scale power (see \Sec{extrapSmallScale}) manifests itself at higher $k\tau$. Hence the $s=0$ case yields lower results at the highest-plotted $k\tau$ values.

Since $s=0$ simulations enable the greatest range in $\tau/\tau'$ under scaling, while accurately matching the dynamics seen at higher $s$ values, our final CMB results will be based upon $s=0$ simulations and we will limit our remaining discussion in this article to simulations at this value of $s$.

\subsection{UETC scaling function results and decoherence}
\label{sec:UETCresults}

\begin{figure*}
\resizebox{0.49\textwidth}{!}{\includegraphics{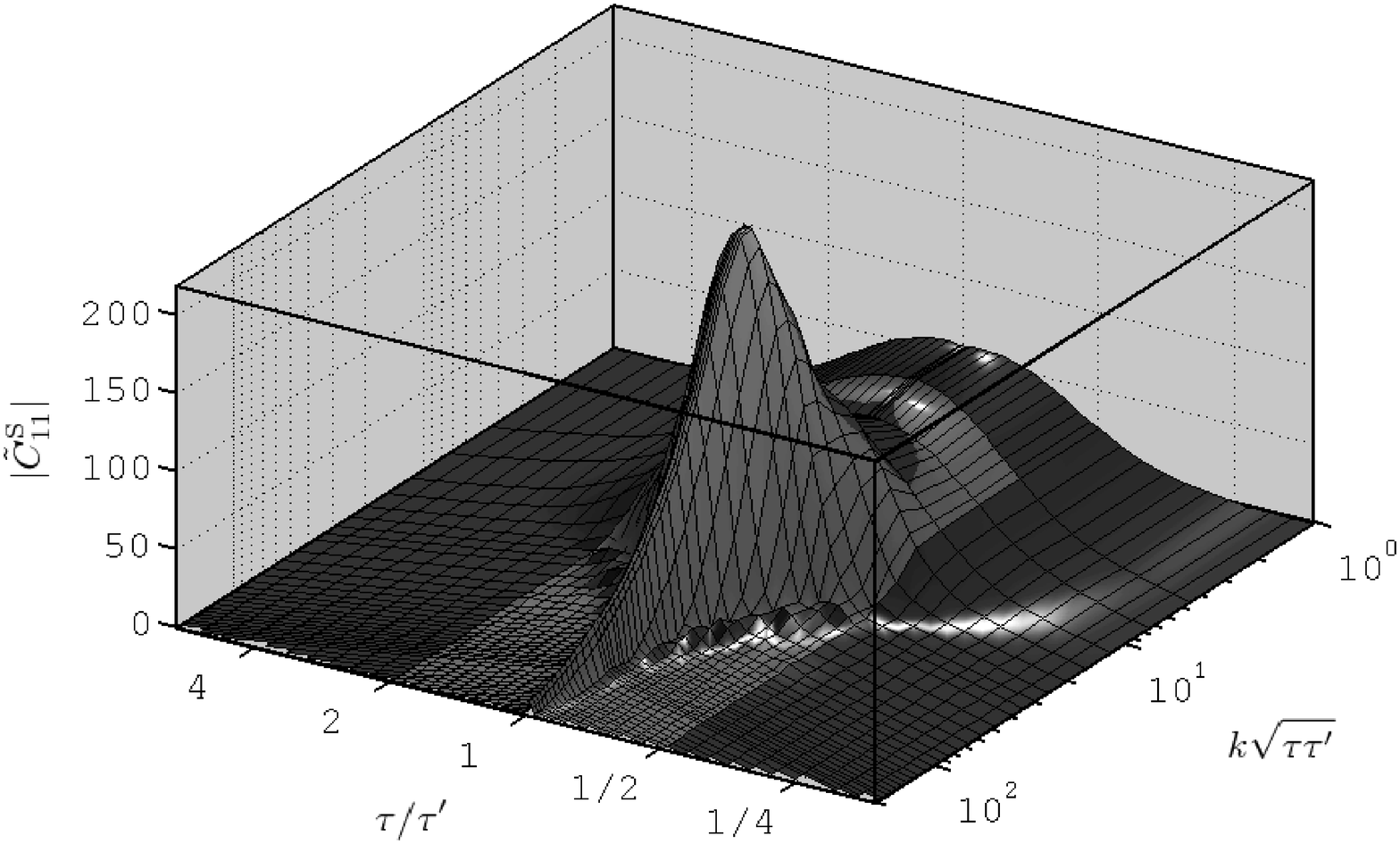}}
\resizebox{0.49\textwidth}{!}{\includegraphics{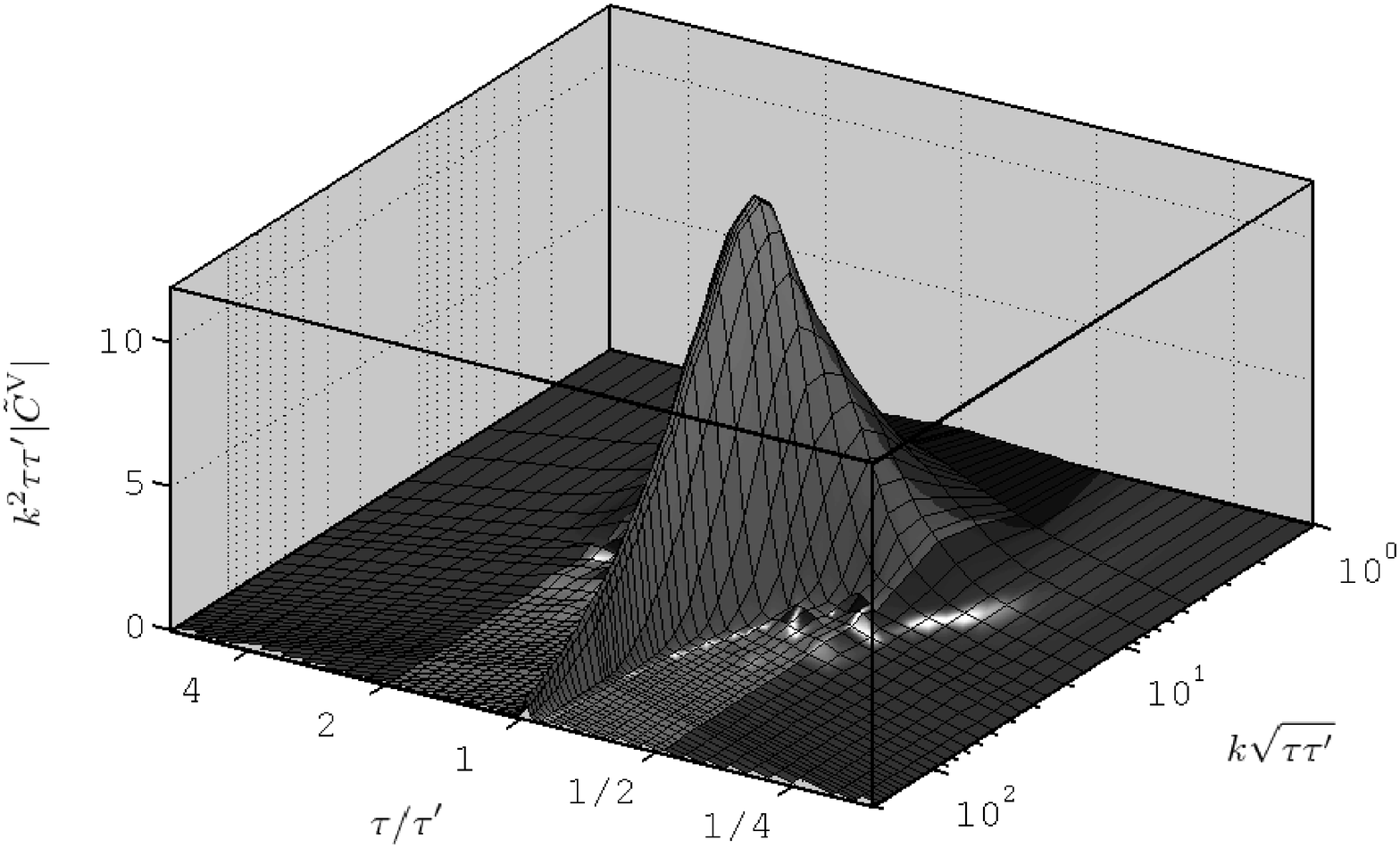}}\\
\resizebox{0.49\textwidth}{!}{\includegraphics{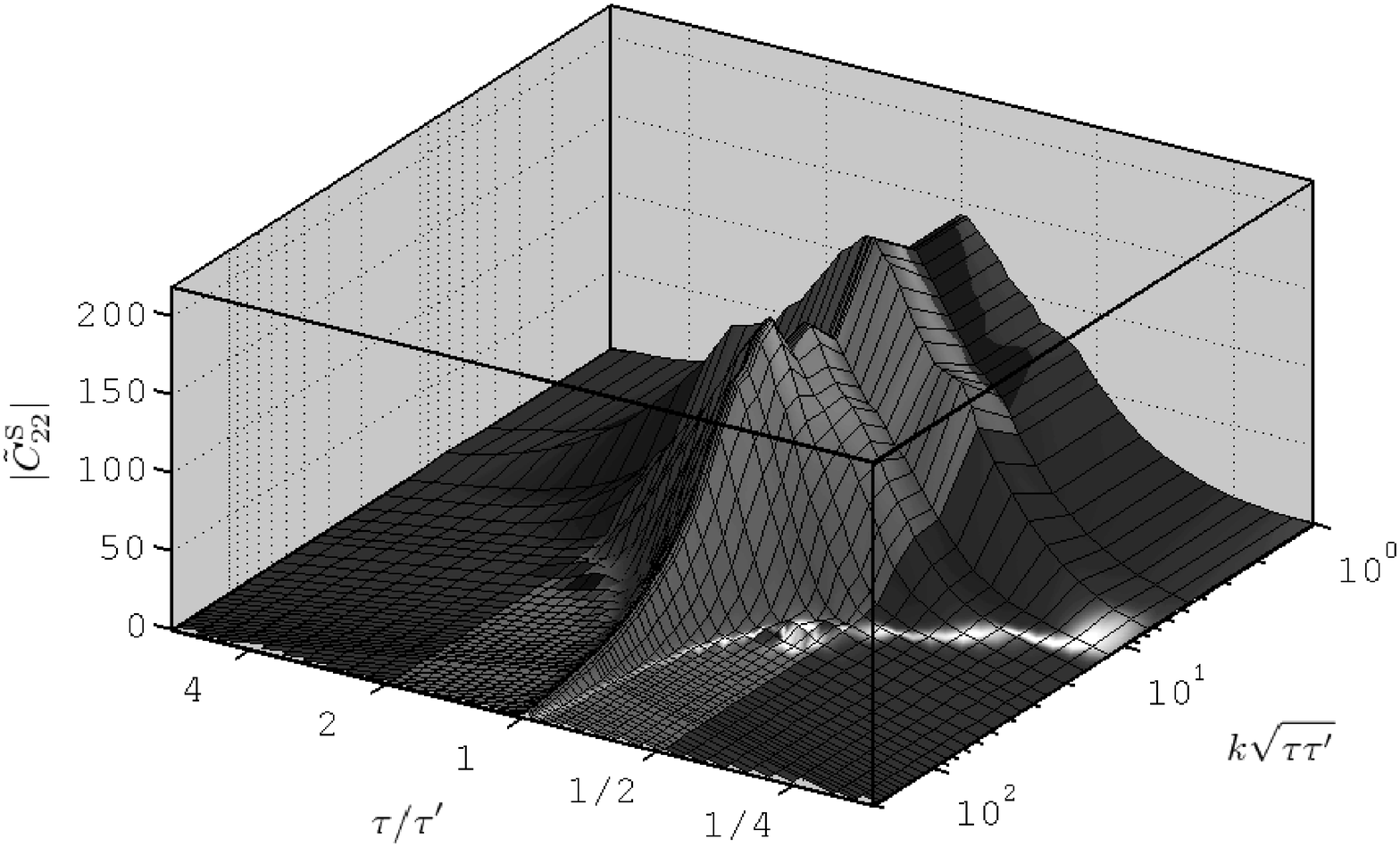}}
\resizebox{0.49\textwidth}{!}{\includegraphics{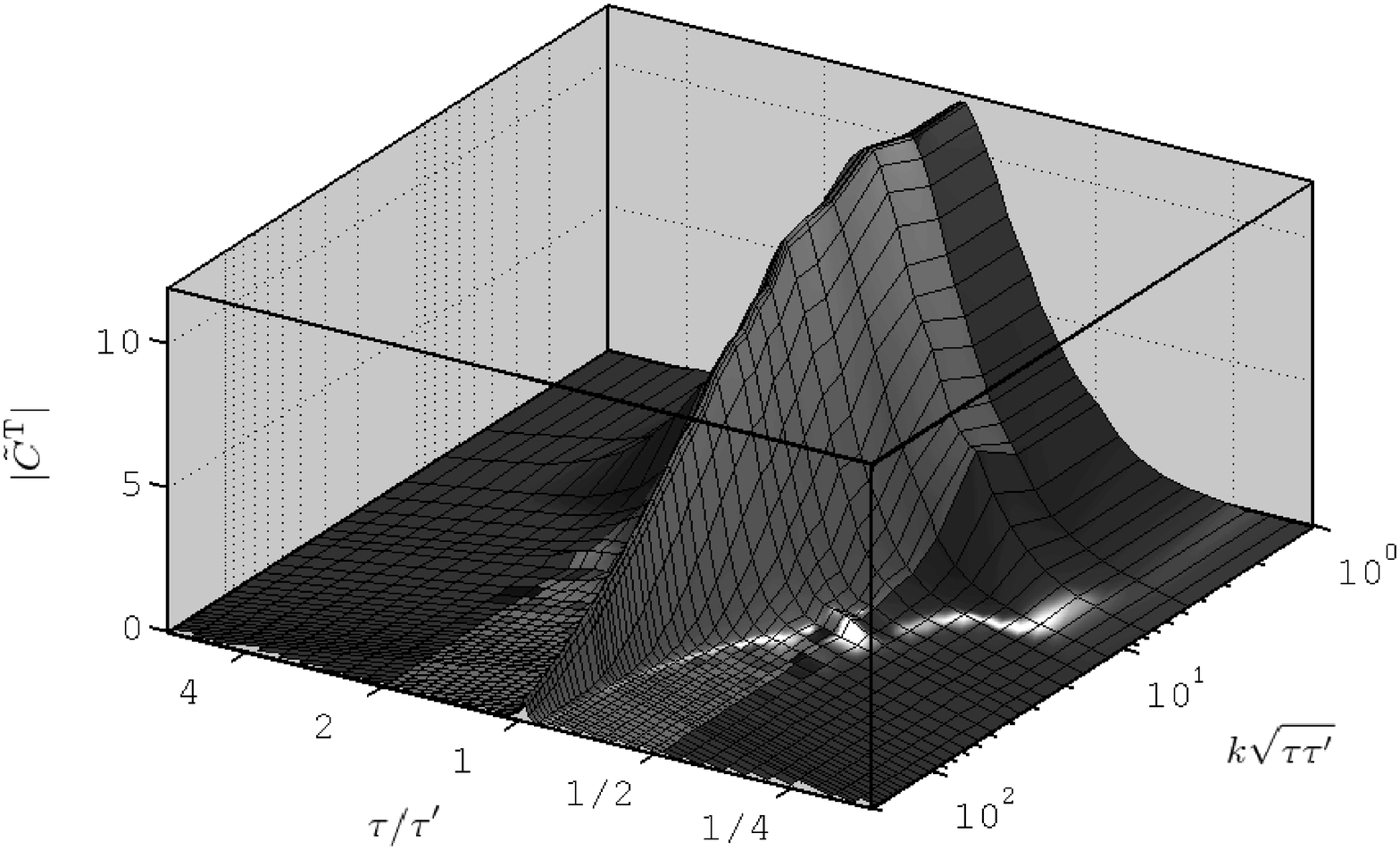}}\\
\begin{flushleft}
\resizebox{0.49\textwidth}{!}{\includegraphics{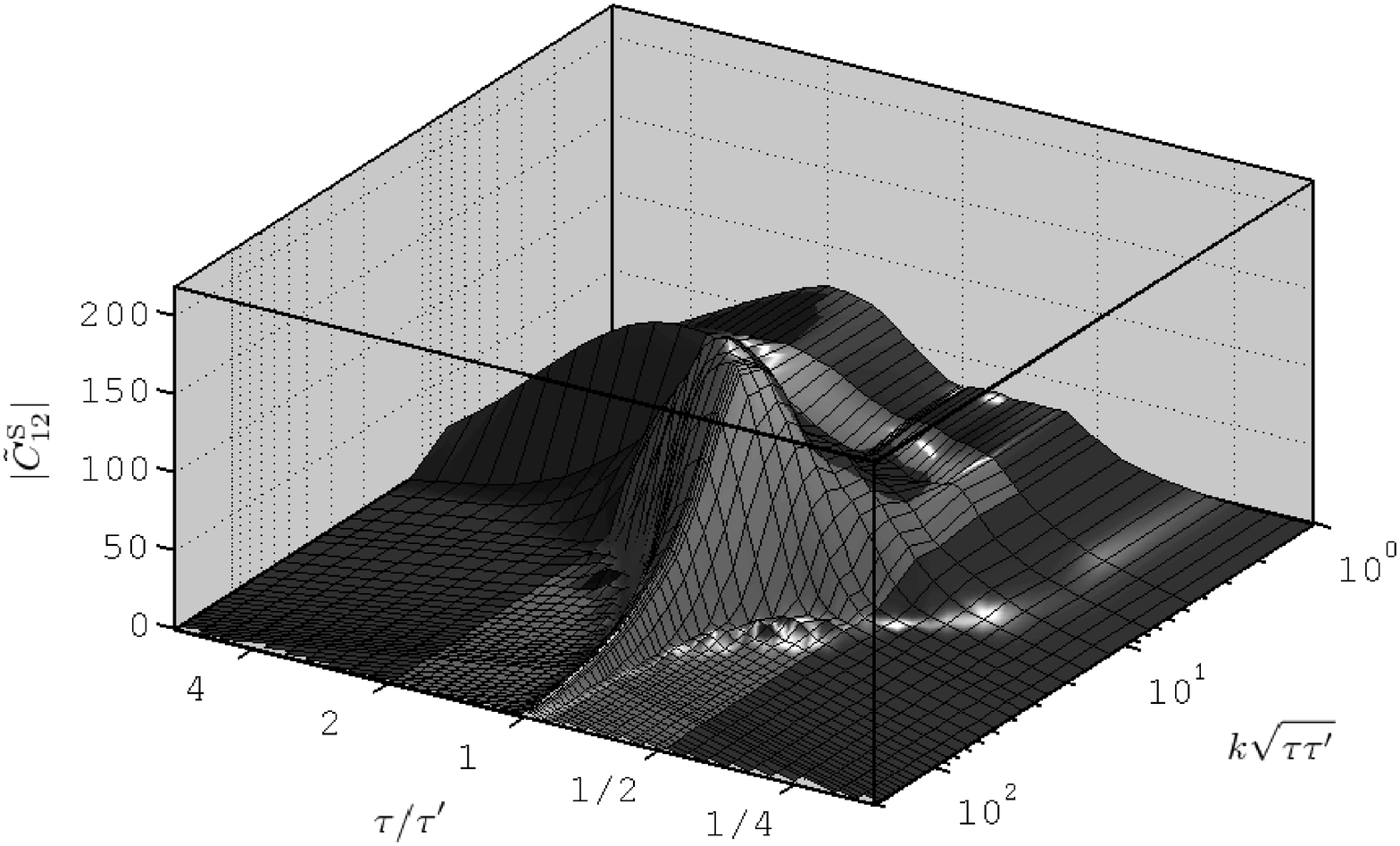}}
\end{flushleft}
\caption{\label{fig:UETCs}The UETC scaling functions $\FT{C}(k\sqrt{\tau\tau},\tau/\tau')$ in the matter era from $1024^{3}$ simulations with $s=0$, averaged over 3 realizations. The raw data is highlighted by the lighter central region, with extrapolations to more extreme $\tau/\tau'$ values and to lower $k\sqrt{\tau\tau'}$ indicated by the darker regions. The vertical axis indicates $|\FT{C}|$ and it should be noted that the cross-correlator is negative near its peak.}
\end{figure*}

Despite our improved initial conditions, we are only able to study the system when it is scaling accurately for conformal time ratios $\tau/\tau'\sim 2$ (corresponding to physical time ratios $\sim 4$ in the radiation era and $\sim 8$ in the matter era). As can be seen in \Fig{UETCs} this is sufficient to map out the important region of the UETCs, and to permit extrapolation in $\tau/\tau'$ as explained below. It can be seen that the auto-correlator scaling functions peak for $\tau=\tau'$ and decay for unequal times, with decay occurring for $\tau/\tau'$ ratios that deviate only slightly from one if $k\sqrt{\tau\tau'}$ is large but more slowly on super-horizon scales. In the cross-correlation case $\FT{C}\Sr_{12}$, this is broadly true but the peak is noticeably offset on super-horizon scales. 

\begin{figure*}
\resizebox{0.49\textwidth}{!}{\includegraphics{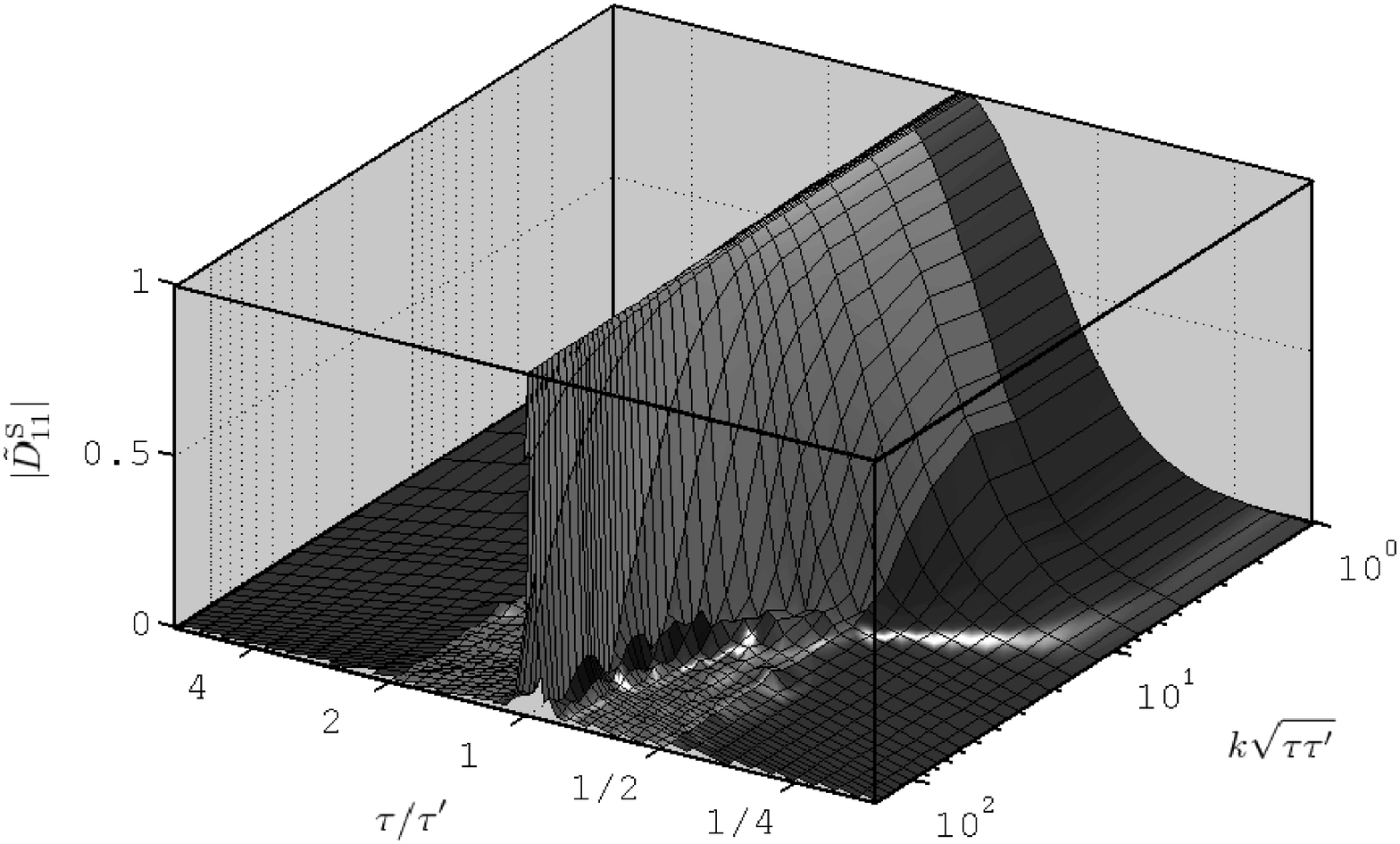}}
\resizebox{0.49\textwidth}{!}{\includegraphics{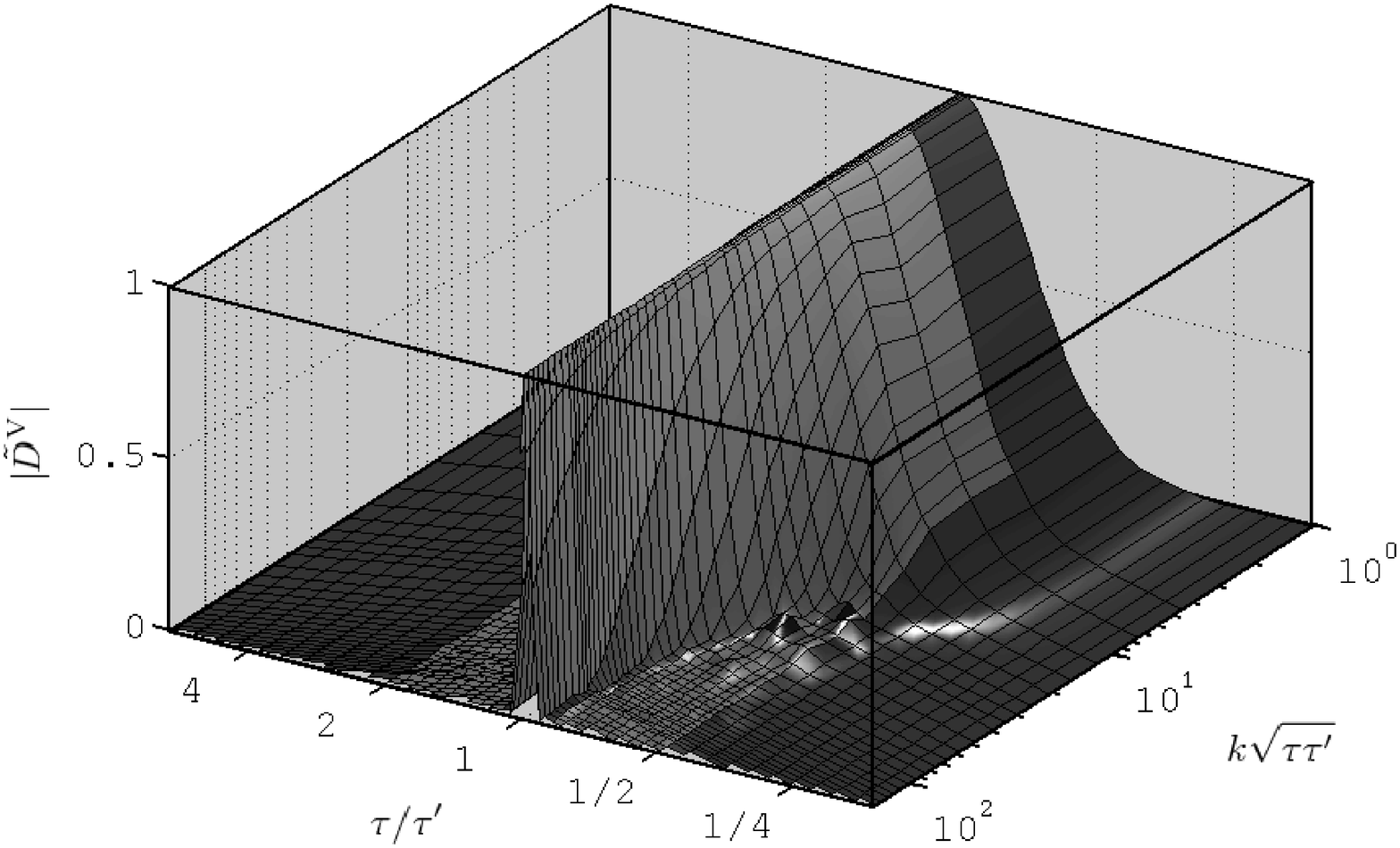}}\\
\resizebox{0.49\textwidth}{!}{\includegraphics{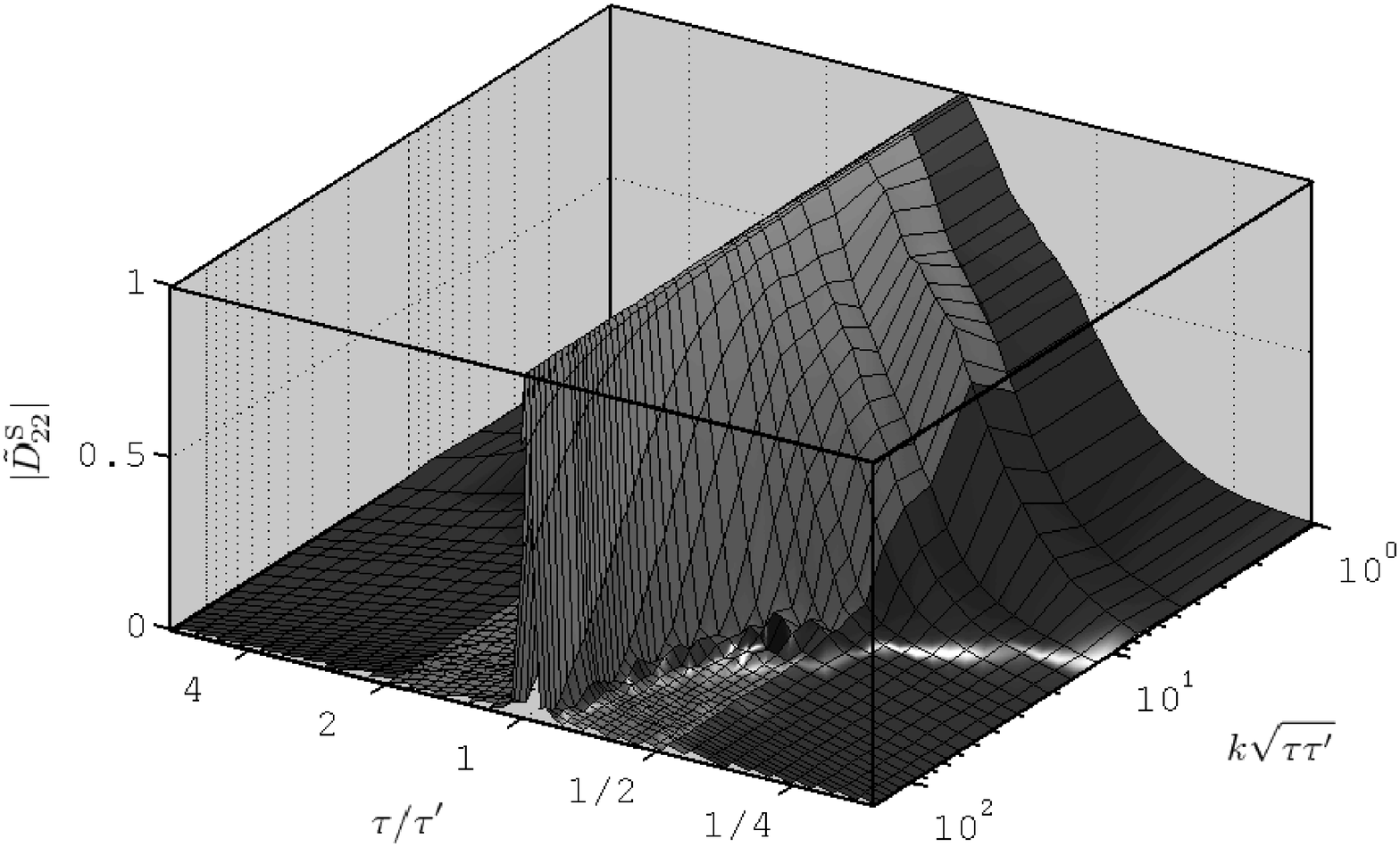}}
\resizebox{0.49\textwidth}{!}{\includegraphics{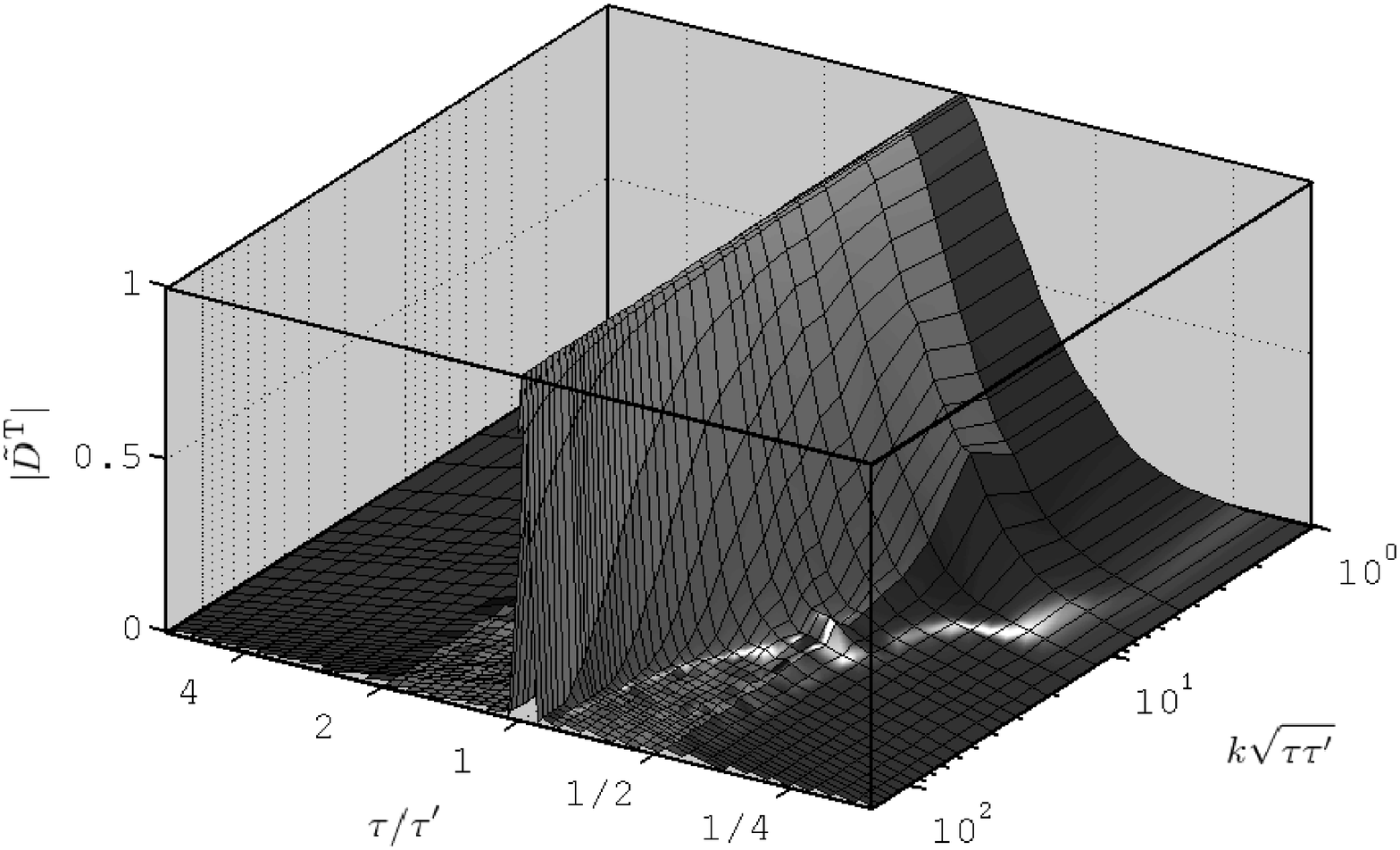}}\\
\begin{flushleft}
\resizebox{0.49\textwidth}{!}{\includegraphics{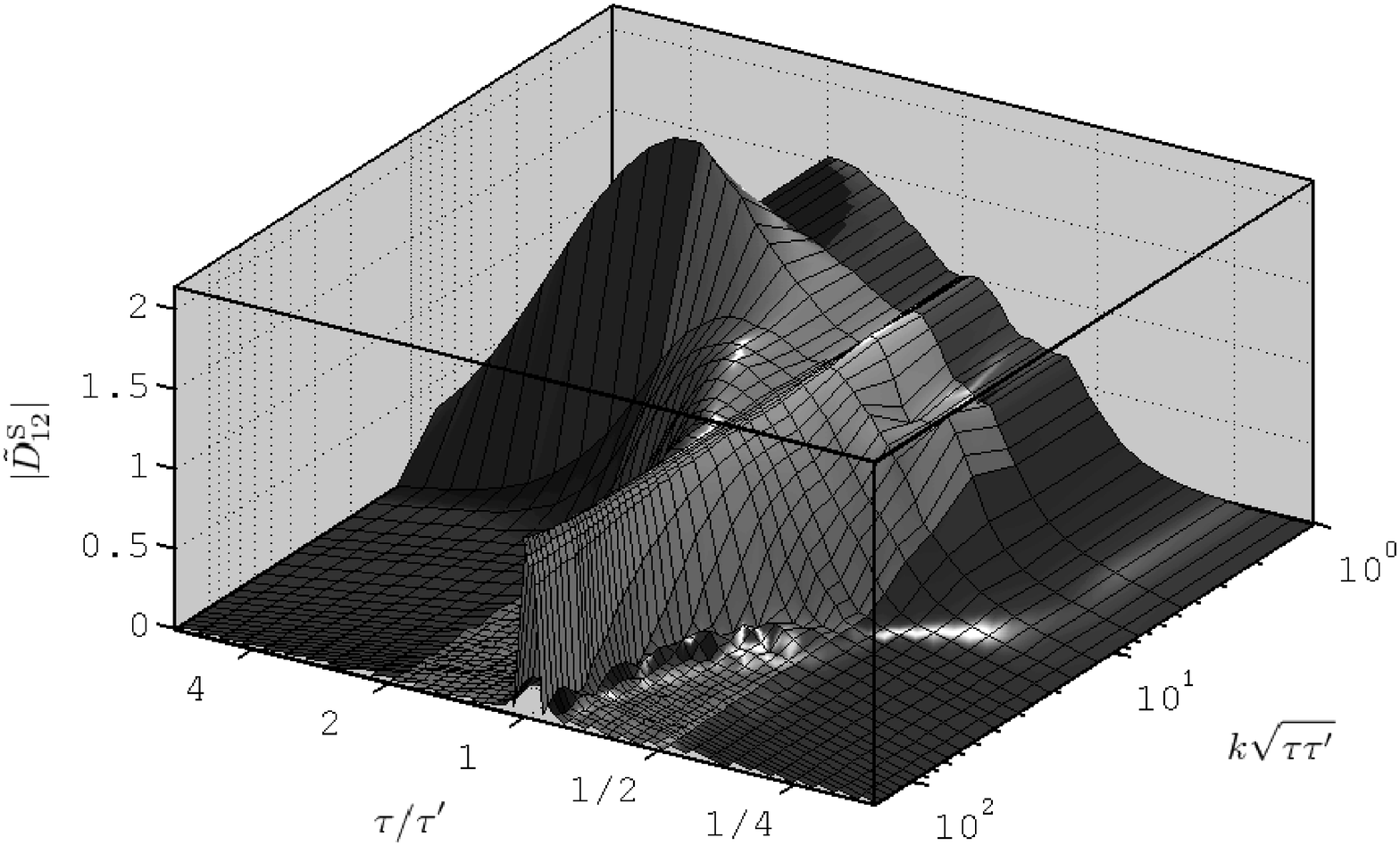}}
\end{flushleft}
\caption{\label{fig:coherence}The decoherence functions $\FT{D}(k\sqrt{\tau\tau},\tau/\tau')$ in the matter era from $1024^{3}$ simulations with $s=0$, averaged over 3 realizations. The raw data is highlighted by the lighter central region, with extrapolations to more extreme $\tau/\tau'$ values and to lower $k\sqrt{\tau\tau'}$ are indicated by the darker regions.}
\end{figure*}

\begin{figure}
\resizebox{\columnwidth}{!}{\includegraphics{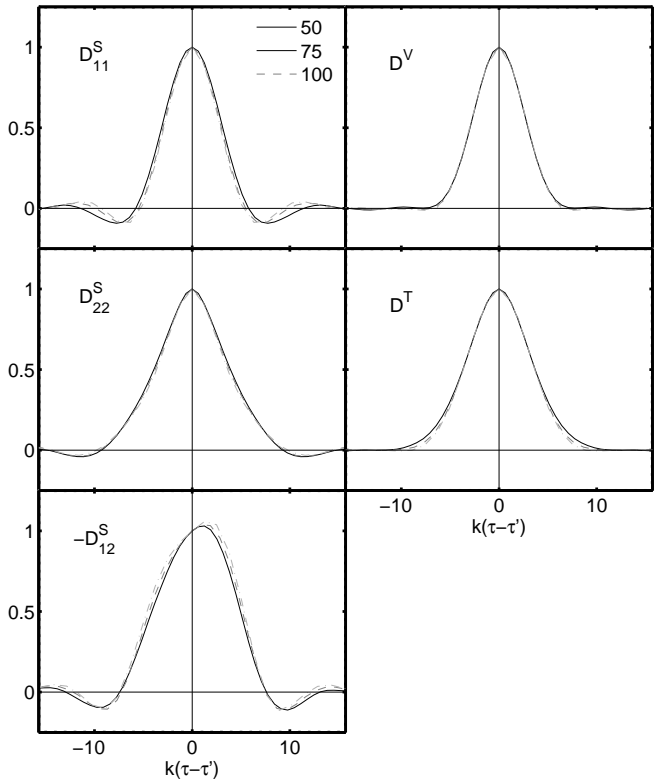}}
\caption{\label{fig:coherenceSlice}Slices through the decoherence functions $\FT{D}$ at small scales for the matter era. In order to demonstrate the approximately constant form at these scales when plotted as a function of $k(\tau-\tau')$, results are shown for the three values $k(\tau+\tau')/2$ indicated by the legend. The latter quantity is perpendicular to $k(\tau-\tau')$ in the $k\tau-k\tau'$ plane (see also \Fig{patches}) and simplifies to $k\tau$ for the equal-time case.}
\end{figure}

This behaviour can be considered in more detail via the coherence function, which we also use for $\tau/\tau'$ extrapolation. We define this function as follows in order to remove the equal-time $k\tau$ dependence from the unequal-time results:
\begin{equation}
\FT{D}(k\sqrt{\tau\tau'},\tau/\tau') 
= 
\frac{\FT{C}(k\sqrt{\tau\tau'},\tau/\tau')}{\sqrt{\left| \FT{C}(k\tau,1) \; \FT{C}(k\tau',1) \right|\;}},
\end{equation}
where the modulus in the square-root is relevant only for $\FT{C}\Sr_{12}$. This has the attractive feature of being equal to $\pm1$ at equal-times and is positive at such times for all auto-correlations. Our results are shown in \Fig{coherence}, \Fig{coherenceSlice}, and also in \Fig{patches}. The small-scale behaviour can be understood by considering that the network can quickly decohere when coarse-grained on sub-horizon scales $1/k \ll \xi$ since the relativistic strings must simply travel $\sim 2\pi/k$. As a result the coherence function decays by \mbox{$|k\tau-k\tau'|\sim2\pi$}. \mbox{Figure \ref{fig:coherenceSlice}} highlights this decay, and that the form of $\FT{D}$ as a function of $k(\tau-\tau')$ is scale-independent on small scales. On the other hand, when plotted as a function of $\tau/\tau'$, as in \Fig{coherence}, the equal-time ridge becomes increasingly sharp on these scales.

On super-horizon scales the decay is much slower, taking of order a Hubble time: $\tau/\tau'\sim2$. For these longer wave modes the coarse-grained regions each contain a number of horizon volumes. A particular horizon volume that is initially over-dense will become under-dense due to the stochastic string dynamics in a time that is of the same order its initial size, i.e. $\sim\tau'$. Therefore the averaged properties of the large coarse-grained region decohere on this time-scale and the decoherence functions for large scales show decay at a fixed value of $\tau/\tau'$, independent of scale.

We extrapolate our UETC scaling function results to greater time ratios in the auto-correlation cases by first noting that the form of the coherence function at fixed $k\sqrt{\tau\tau'}$ is approximately Gaussian in $\log(\tau/\tau')$, but with the width dependent on $k\sqrt{\tau\tau'}$. For super- and near-horizon scales we hence take each $k\sqrt{\tau\tau'}$ value in turn and match a Gaussian profile $\exp[-\log^2(\tau/\tau') / 2\sigma^2]$ to the data at the most extreme $\tau/\tau'$ value and then use that profile to extrapolate to larger time ratios. On the other hand, for sub-horizon scales the decay is very rapid and we simply extrapolate with zeros. The results of these extrapolations can be seen in Figs. \ref{fig:UETCs} and \ref{fig:coherence}.

For the cross-correlator $\FT{C}\Sr_{12}$ on the other hand, our results show that the $\tau/\tau'$-profile on horizon and super-horizon scales is complex and non-Gaussian, making reliable extrapolation impossible. For this correlator we use the Gaussian fits to $\FT{C}\Sr_{11}$ and $\FT{C}\Sr_{22}$, with widths $\sigma_{11}$ and $\sigma_{22}$, to provide information on the likely decay timescale. We then fit a Gaussian of width $\sigma_{12}=(\sigma_{11}+\sigma_{22})/2$ but with free mean and normalization to the $\FT{C}\Sr_{12}$ data from the most extreme time ratios, and then employ that to yield extrapolation. 


\subsection{Extrapolation of UETCs to sub-string scales}
\label{sec:extrapSmallScale}

\begin{figure}
\resizebox{\columnwidth}{!}{\includegraphics{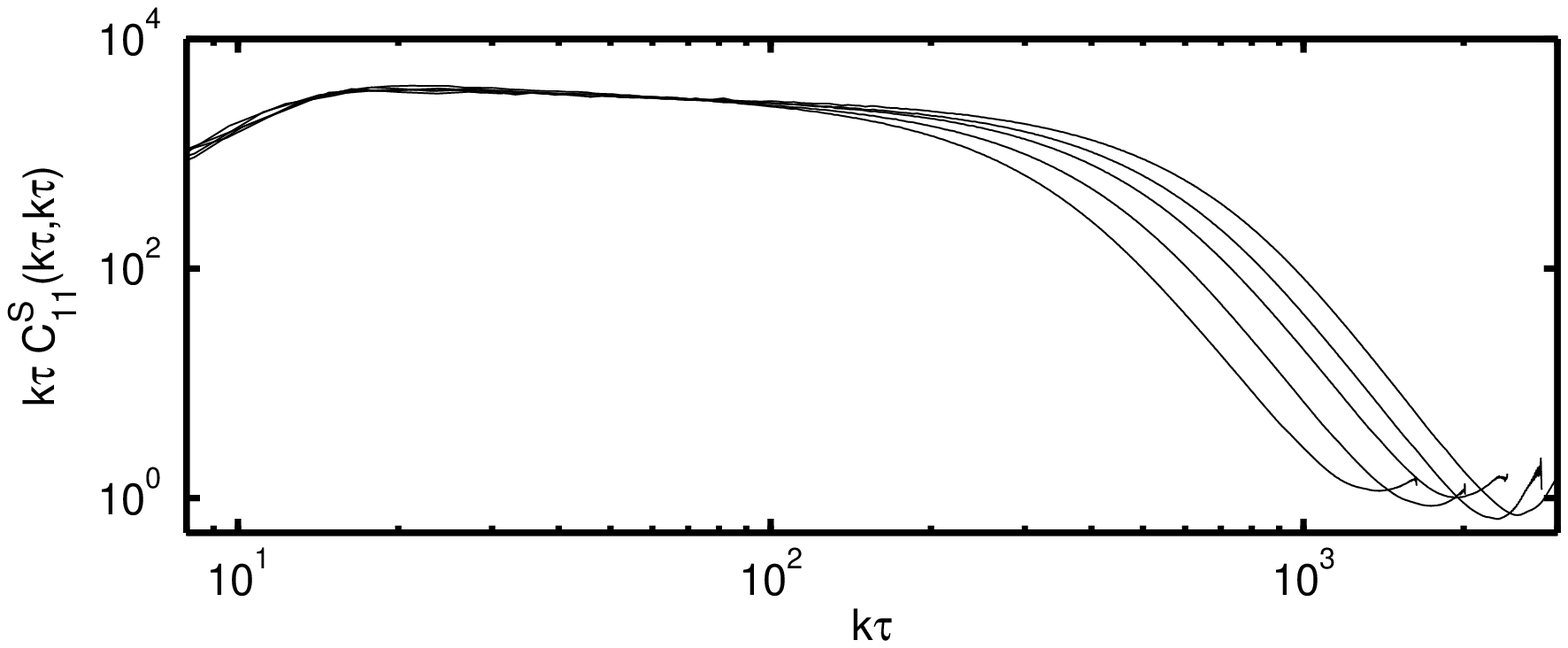}}
\resizebox{\columnwidth}{!}{\includegraphics{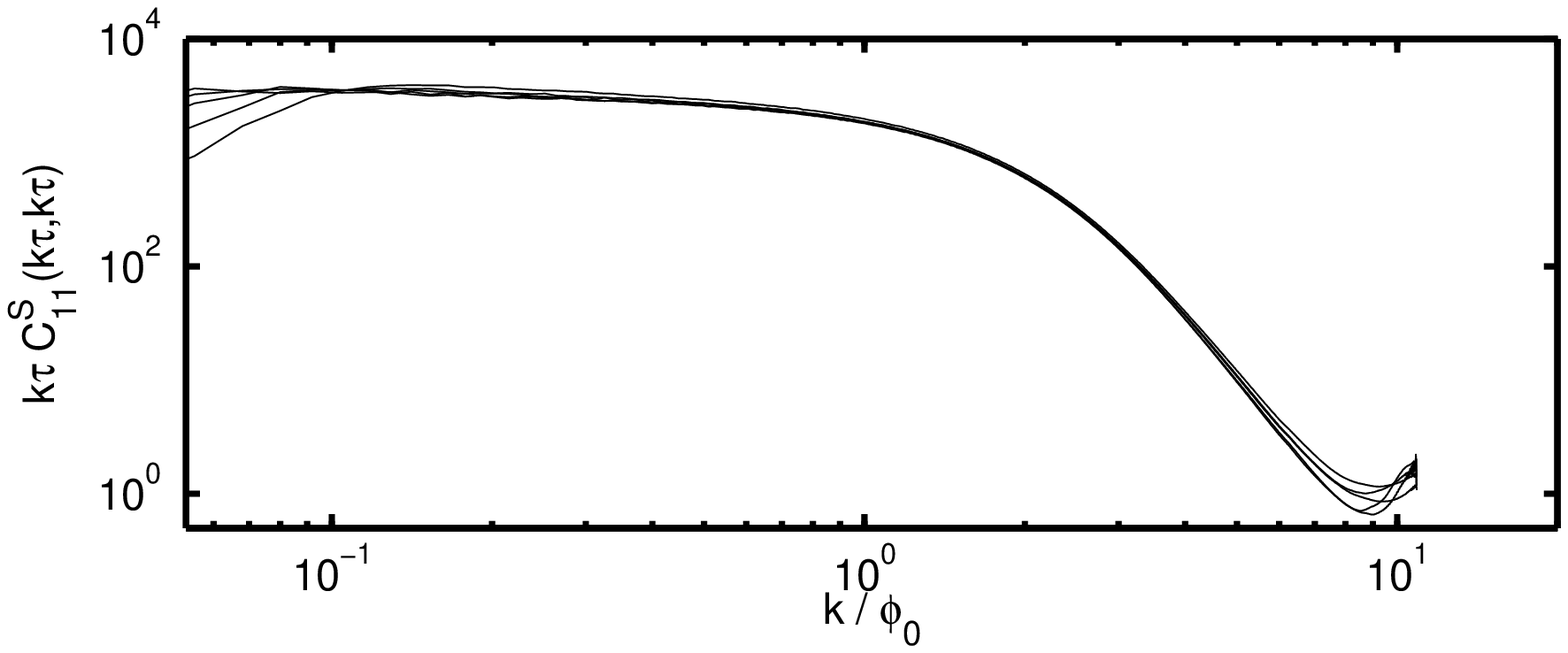}}
\caption{\label{fig:ETCdropOff}Results for $k\tau \FT{C}\Sr_{11}$ from $s=0$ simulations, plotted against $k\tau$ (upper pane) and $k$ (lower pane). The 5 lines correspond to equally spaced times between $\tau=150$ and $300 \VEV^{-1}$. In the upper pane the 5 lines are indistinguishable at low $k\tau$ due to the observed scaling, but scaling is broken on small scales where lines progressively move to the right. In the lower pane the lines are indistinguishable at high $k$, showing that on such scales $k\tau \FT{C}\Sr_{11}$ is then simply a function of $k$, ie. the comoving length scale, and indicating that the attenuation of the signal occurs on scales a few times the comoving string width, which is constant in this $s=0$ case.}
\end{figure}

Figure \ref{fig:ETCdropOff} shows a plot of $k\tau \FT{C}\Sr_{11}$ for $\tau'=\tau$ from simulations with $s=0$. From the upper pane it can be seen that this quantity is approximately independent of $k\tau$ on sub-horizon scales until $k\tau\sim200$ at the earliest time shown or until about $k\tau\sim400$ at the latest time plotted. The plateau highlights an important point: that $\FT{C}\Sr_{11}$ drops off as $\approx1/k\tau$ on sub-horizon scales, which matches our basic expectations as well as similar measures from NG simulations \cite{Vincent:1996qr}. However, on smaller scales there is a sudden attenuation of power, and at a $k\tau$ value that is increasing with time. The lower pane clarifies the later by re-plotting this against $k$ rather than $k\tau$, showing that this is occurring at a fixed comoving scale of $k\approx2 \VEV$. This corresponds to a few times the string width, which is a fixed comoving width in this $s=0$ simulation. 

\begin{figure}
\resizebox{\columnwidth}{!}{\includegraphics{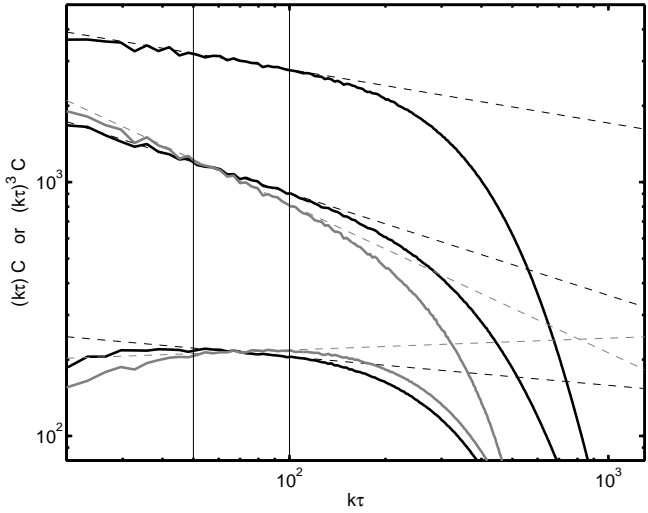}}
\caption{\label{fig:ETCextrap}Results for all five equal-time scaling functions multiplied by $k\tau$ (or $(k\tau)^{3}$ in the vector case), with power-law fits over the range $k\tau=50\rightarrow100$ (between the two vertical lines). The uppermost line is for $\FT{C}\Sr_{11}$, then of the middle pair of lines the black one is $\FT{C}\Sr_{22}$ and the gray one is $|\FT{C}\Sr_{12}|$, and finally of the lower pair the black one is $\FT{C}\Vr{}$ and the gray one if $\FT{C}\Tr{}$. A horizontal fit line would indicate that $\FT{C}$ (or $(k\tau)^{2}\FT{C}\Vr{}$ in the vector case) varies as $1/k\tau$.}
\end{figure}

We find evidence that all equal-time scaling functions vary as approximately $1/k\tau$ on small scales (except in the vector case for which $(k\tau)^2\FT{C}\Vr{}$ varies roughly as $1/k\tau$). Figure \ref{fig:ETCextrap} shows power-law fits to $k\tau\FT{C}$ (and $(k\tau)^3\FT{C}$ in the vector case) over the range $k\tau=30\rightarrow100$, which lies between the interesting effects on horizon-scales and the artifacts of the string width on very small-scales. While it would be desirable to have a further order of magnitude via which to confirm the behavior in this regime, it is clear in all cases that a substantial improvement in our estimate of the scaling functions on small scales would be arrived at by extrapolating this trend down to scales near and below the string width.

We extrapolate the unequal time scaling functions to small scales using our knowledge of the attenuation experienced at equal-times for both $\tau$ and $\tau'$. First we define the attenuation level $R$ as the ratio of the extrapolated equal-time scaling function and the measured version at time $\tau$:
\begin{equation}
R(k,\tau) = \frac{Q\;(k\tau)^{-p}}{\FT{C}_{\tau}(k\tau,1)},
\end{equation}
where we have added a subscript to $\FT{C}$ to indicate that the measured correlator does not scale exactly and so is not only dependent upon $k\tau$, while $p$ and $Q$ are constants found from the power-law fit. We then extrapolate the UETC scaling function as:
\begin{equation}
\FT{C}(k\sqrt{\tau\tau'},\tau/\tau') = \sqrt{R(k,\tau)R(k,\tau')} \; \FT{C}_{\tau}.
\end{equation}
This is based on the correlators being quadratic quantities and includes an appropriate compensation factor for each of the two times involved, which in practice must be nearly equal for the scaling function to be significant, while in the equal-time case this simply returns the power-law fit. 


\subsection{CMB calculations and UETC eigenfunction decomposition for small scales}
\label{sec:eigenDecomp}

As explained in \Sec{method}, having determined the UETC scaling functions $\FT{C}(k\sqrt{\tau\tau'},\tau/\tau')$, we decompose them into a sum of terms involving functions of a single variable $\FT{c}_{n}(k\tau)$, which then act as the sources of metric perturbations in the CMB calculations. Each function hence corresponds to one term in the sum for the CMB temperature or polarization power spectra. 

The following discussion is simplified if we discuss initially only the tensor case, for which we require a single UETC that is symmetric under the exchange of $\tau$ and $\tau'$. Since we can only represent $\FT{c}_{n}(k\tau)$ numerically at discrete $k\tau$ values $\kappa_{i}$ (with $i=1,\ldots,M$), then this step boils down to the decomposition of a real and symmetric $M \times M$ matrix $\FT{C}_{ij} = \FT{C}(\sqrt{\kappa_i \kappa_j},\kappa_i/\kappa_j)$, the eigenvectors of which form an orthonormal set. It can hence be seen that the decomposition:
\begin{equation}
\FT{C}_{ij} = \sum_{n=1}^{M} \lambda_{n} \FT{c}_{ni} \FT{c}_{nj},
\end{equation}
is equivalent to determining the eigenvalues $\lambda_{n}$ and eigenvectors $\FT{c}_{ni}$ of the matrix $\FT{C}_{ij}$. In the scalar case, the situation is more complex but the discussion proceeds similarly\footnote{In the scalar case there are two source functions $\FT{S}_{\Phi}$ and $\FT{S}_{\Psi}$ between which there is a finite correlation. To deal with this we form a $2M \times 2M$ symmetric matrix by tiling $\FT{C}\Sr_{11}$, $\FT{C}\Sr_{12}$, $\FT{C}\Sr_{21}$ and $\FT{C}\Sr_{22}$ and then take $\FT{S}_{\Phi}$ to be the first $M$ elements of the eigenvector and $\FT{S}_{\Psi}$ to be the second half (see BHKU). However, this barely changes the present discussion of small scales}, while in the vector case we apply the decomposition to $k^2\sqrt{\tau\tau'}\FT{C}$. 

In BHKU we found that accurate results over the relevant angular scales were obtained with $M=512$ and $\kappa_{M}=200$, while only the terms with the $128$ highest $|\lambda_n|$ values were needed. However, here we wish to include the high-$k\tau$ tails of the UETCs out to $k\tau\sim5000$, but the narrow width of the equal-time ridge requires $\Delta\kappa\approx1$ and hence we would require $M\sim5000$. Additionally, the small amplitude of the tails implies that their signal is likely to be contained in eigenvectors with very low eigenvalues, and therefore we would need to include all terms in our CMB calculations. Due to the nature of the sources, each Einstein-Boltzmann integration is much slower than the corresponding primordially-seeded calculation required for CMB predictions from inflation and hence this process would be particularly time consuming. Further, the contributions to the CMB power spectrum in our target range $2\le\ell\le4000$ from extremely high $k\tau$ are minor, while our knowledge of the UETCs on such scales is only via the above power-law extrapolations.

We proceed instead by performing the decomposition in such way that each ``eigenvector'' is localized in $k\tau$, since we then have an immediate understanding of how it contributes to the CMB power spectra. Firstly we arrange for all of the dominant horizon and super-horizon power ($k\tau\lesssim100$) to be contained within a particular set of eigenvectors, which then completely dominate the CMB temperature power spectrum for multipoles $\ell\lesssim1000$, but their contributions decay for smaller angular scales. Secondly, the simple form of the UETC scaling functions on sub-horizon scales allows us to obtain a knowledge of the CMB contributions from extremely high $k\tau$ values by combining our calculations for moderate $k\tau$ values with approximate scaling laws, as explained momentarily.

\begin{figure}
\resizebox{\columnwidth}{!}{\includegraphics{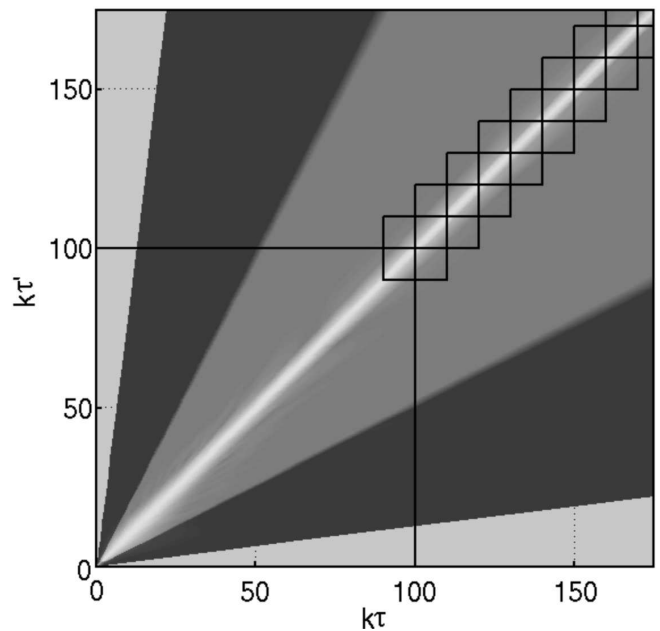}}
\caption{\label{fig:patches}The block-wise coverage of the $k\tau-k\tau'$ plane. A large primary block contains the dominant UETC data at near- and super-horizon scales ($k\tau<100$). Then smaller blocks of side $\Delta(k\tau)=20$ cover the diagonal, with overlapping to aid coverage. The gray-scale image shows the coherence function $\FT{D}\Sr_{11}$ to illustrate that these patches cover the region over which the UETC is significant, with light colors indicating higher values. The dark outer region indicates the region for which we extrapolate the coherence function, as in \Fig{coherence}.}
\end{figure}

The $k\tau$ localization is achieved by noting that the rapid decay of the scaling functions for unequal times means that $\FT{C}_{ij}$ can be approximated as the sum:
\begin{equation}
\FT{C}_{ij} \approx \sum_{m} \FT{C}_{ij}^{m}, 
\end{equation}
where, as indicated in \Fig{patches}, $\FT{C}_{ij}^{m}$ is finite only for $\delta_{m} \le i,j < \delta_{m+2}$, with the $\delta_{n}$ chosen to yield an array of overlapping matrix blocks which cover the important regions of the $k\tau$-$k\tau'$ plane: horizon scales and the sub-horizon equal-time ridge. In the regions of overlap, the UETC power is shared between the two matrix blocks such that the contribution from a given block varies smoothly from zero at the extremes of the corresponding $k\tau$ range, up to full at the block centre. As a result of this construction, each component matrix $C_{ij}^m$ has eigenvectors $\FT{c}_{i}^{mn}$ which have the desired locality in $k\tau$, being finite only for $\delta_m \le i < \delta_{m+2}$, while there are $\delta_{m+2}-\delta_m$ eigenvectors for each block. We chose the $\delta_m$ values and $k\tau$ spacing such that the first block is finite only for $k\tau<100$, while the subsequent blocks have width $\Delta(k\tau)=20$, as indicated in \Fig{patches}. 

Importantly, the content in all of the higher blocks is of the same form: a ridge of given width with a central height that decays by a fraction $\sim \Delta k\tau / k\tau$ across the block (in addition to the decays required to share power between overlapping sub-matrices). So long as $\Delta k\tau << k\tau$, then the $\sim 1/k\tau$ power-law decay has minimal impact and the differences between the higher blocks are effectively just translation in $k\tau$ and a change in normalization. That is, the eigenvectors are effectively translated in $k\tau$ while the eigenvalues absorb the normalization change.

\begin{figure}
\resizebox{\columnwidth}{!}{\includegraphics{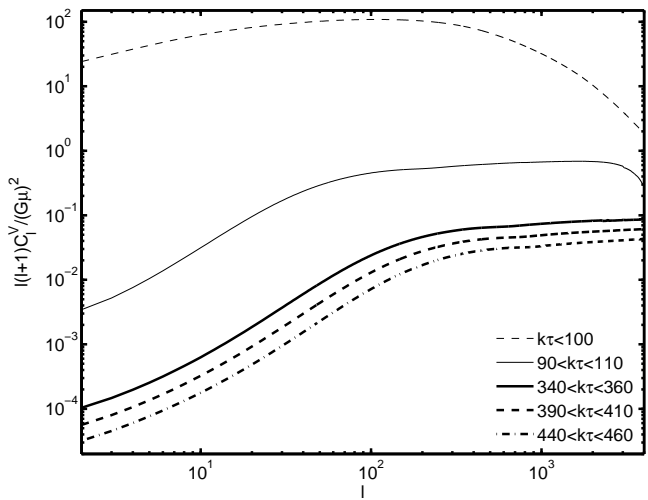}}
\caption{\label{fig:patchScaling}Contributions to the vector component of the temperature power spectrum from discrete $k\tau$ ranges. The highest line shows the dominant contribution from horizon and super-horizon scales. The contribution from first sub-horizon matrix block is shown in the middle of the plot, while the approximate scaling law is indicated by the high $k\tau$ lines, which are lowest in the figure.}
\end{figure}

\begin{figure}
\resizebox{\columnwidth}{!}{\includegraphics{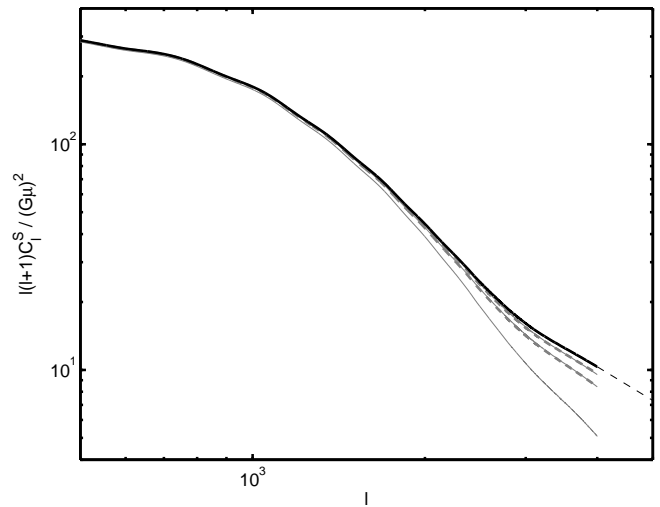}}
\caption{\label{fig:patchSum}The build up of the high $\ell$ scalar mode temperature power spectrum as the maximum included $k\tau$ value is increased. From bottom to top in the plot these correspond to the $k\tau\lesssim150$, $k\tau\lesssim300$, $k\tau\lesssim550$ and finally $k\tau\lesssim5000$. Results shown by thin solid grey lines are from full CMB calculations, while the thick grey dashed and thick black solid lines correspond to results obtained using the rescaling property for $k\tau\gtrsim150$. The power-law extrapolation is shown by the thin dashed line and is a reasonably accurate description of the final $k\tau\lesssim5000$ result for $\ell\gtrsim3000$.}
\end{figure}

When our modified version of CMBEASY is applied to the eigenvalues and eigenvectors from the higher blocks, the temperature and polarization power spectra returned also have effectively the same form, except for the approximate re-scaling:
\begin{equation}
\label{eqn:patchScale}
\mathcal{P}_{\ell}^{m} = \ell(\ell+1) \mathcal{C}_{\ell}^{m} \approx \frac{f(\ell/\kappa_{m})}{\kappa_{m}^\beta},
\end{equation}
where $\kappa_{m}$ is the central $k\tau$ value of the $m$th block while $\beta\approx2$ is a constant to be determined. This is illustrated for the vector mode in \Fig{patchScaling}. It can also be seen that the contribution from a given block rises as approximately $\propto\ell$ up to a plateau for $\ell\approx\kappa_{m}\rightarrow200\kappa_{m}$ before rapidly decaying for higher $\ell$. The key point is that given this scaling, we can find an approximation to the CMB power spectrum contribution from $k\tau$ values beyond the maximum value for which we perform full calculations $\kappa_\mathrm{full}$, by simply summing over many such re-scaled forms until the desired accuracy is reached. The total power spectrum is hence:
\begin{equation}
\label{eqn:patchSum}
\mathcal{P}_{\ell} 
= 
\sum_{m=1}^{m_\mathrm{full}} \mathcal{P}_{\ell}^{m} \;\;\;
+ 
\sum_{m=m_\mathrm{full}+1}^{m_\mathrm{max}} \frac{f(\ell/\kappa_{m})}{\kappa_{m}^{\beta}},
\end{equation}
which is shown in \Fig{patchSum} for the scalar mode. Furthermore, since $\mathcal{P}_{\ell}^{m}$ decays rapidly for $\ell \gtrsim 200\kappa_{m}$, then we may extrapolate the sum for $\ell\rightarrow\infty$ with $m_\mathrm{max}\rightarrow\infty$ by approximating it as an integral, which then varies as:
\begin{equation}
\mathcal{P}_{\ell} 
\propto
\frac{1}{\ell^{\beta-1}},
\end{equation}
independent of the detailed form of the function $f(\ell)$. For $\beta\approx2$, this yields roughly the $1/\ell$ form expected at very small angular scales due to the GKS effect \cite{Hindmarsh:1993pu}, as will be discussed in \Sec{compNG}. For the scalar, vector and tensor modes respectively we find $\beta=2.5$, $1.9$ and $1.7$ and we hence tentatively predict that the CMB contributions from each vary as $\ell^{-1.5}$, $\ell^{-0.9}$, $\ell^{-0.7}$ at very high $\ell$ (see next section). 

Note that the use of this approximate re-scaling property is applied only to $k\tau$ values for which our knowledge of the UETC scaling functions is arrived at via power-law extrapolation. That is, we perform full CMB calculations for the range of $k\tau$ for which the simulations provide direct information, and it is fortunately the case that these provide the overwhelmingly dominant contribution for scales $\ell\lesssim1500$, while in the present article we are limiting ourselves to $\ell\le4000$ and hence the extrapolations made are fairly reliable. 


\section{CMB Results}
\label{sec:CMBresults}


\subsection{Temperature Power Spectrum}

\subsubsection{Results}

\begin{figure}
\resizebox{\columnwidth}{!}{\includegraphics{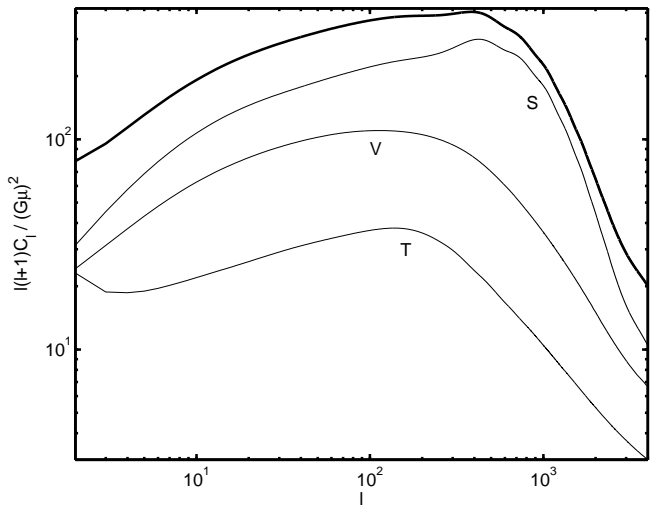}}
\caption{\label{fig:CMBtotal}The CMB temperature power spectrum determined from $s=0$ simulations and incorporating estimated UETC power for $k\tau\lesssim5000$. The plot shows the total (thick line) plus the decomposition (thin lines) into scalar (S), vector (V) and tensor (T) modes.}
\end{figure}

\begin{figure}
\resizebox{\columnwidth}{!}{\includegraphics{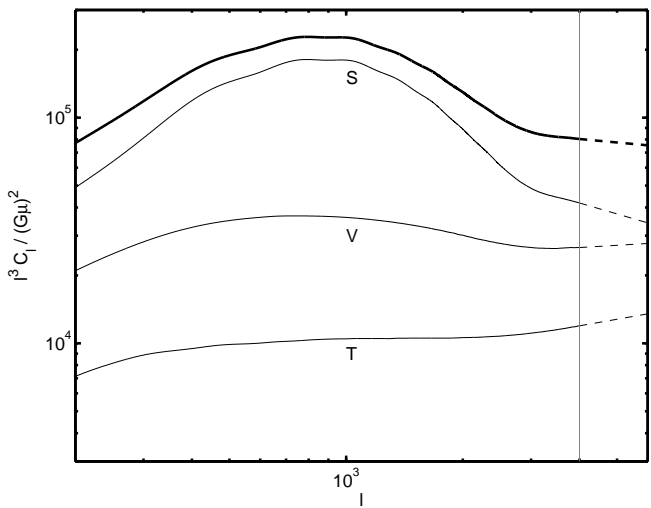}}
\caption{\label{fig:CMBextrap}Results for $\ell^{3}\mathcal{C}_\ell$, highlighting the form of the power spectrum at high $\ell$. The plot shows the total (thick line) plus the decomposition (thin lines) into scalar (S), vector (V) and tensor (T) modes. Additionally, the right of the figure shows the results of our tentative power-law extrapolations to $\ell>4000$.}
\end{figure}

We present our final CMB power spectrum results for cosmological parameters: $h=0.72$, $\Obhh=0.214$ and $\Ol=0.75$, which match those used in BHKU and the central values from non-CMB measurements \cite{Freedman:2000cf,Kirkman:2003uv,Knop:2003iy}. We additionally assume an optical depth to last-scattering of $\optdepth=0.1$, which again matches BHKU. For the case of the temperature power spectrum these are shown in \Fig{CMBtotal}.
 The form is essentially that found in BHKU: at low $\ell$ there is a roughly $\propto\log\ell$ rise up to a broad peak between $\ell\approx30\rightarrow700$ (75\% of peak), with the peak itself at $\ell\approx400$. At greater $\ell$ the spectrum decays, initially as roughly $1/\ell^{2}$, but this then slows to roughly $1/\ell$ for $\ell\gtrsim3000$, which is more clearly evident in \Fig{CMBextrap}. 

\begin{figure}
\resizebox{\columnwidth}{!}{\includegraphics{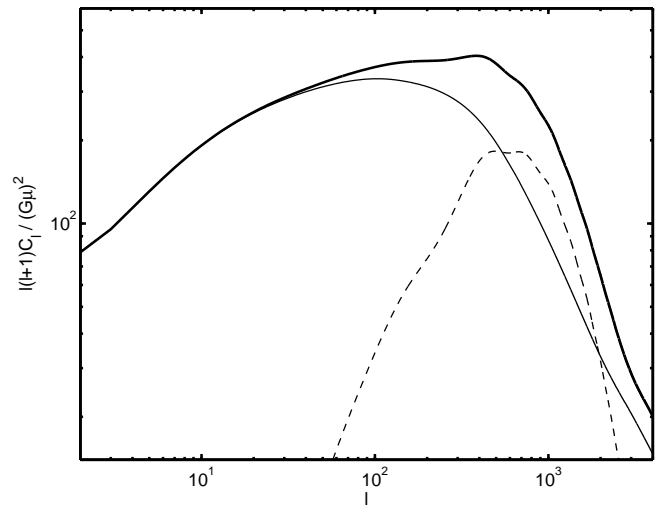}}
\caption{\label{fig:CMBpLSo}The string contribution to the temperature power spectrum when including the string sources only after recombination (thin) or at all times (thick), with the difference between the two additionally shown (dashed). 
Recombination is not instantaneous and additionally the string source functions must be temporally differentiable and thus must be gradually switched on, hence this decomposition is ambiguous and there is an artificial reduction in the amplitude of the power-law at high $\ell$ for the post-recombination results.}
\end{figure}

The reason for this change in behaviour can be seen in \Fig{CMBpLSo}, in which we compare our results to those in which the cosmic string sources are artificially zeroed for times prior to recombination. This zeroing can be seen to remove from the signal a contribution that dominates for $\ell\approx500\rightarrow2000$ over a smaller signal that is sourced at later times and that has the approximate $1/\ell$ form. That is, the early-time signal is Silk-damped at small-scales such that the late-time signal is revealed and dominates for $\ell\gtrsim3000$. Note that this decomposition into pre- and post-recombination is ambiguous since decoupling is not instantaneous and additionally we require the scalar and vector sources to be temporally differentiable, implying that we must have a gradual transition from zero to the nominal source value. These are responsible for the small difference in amplitude between the two power-laws seen in the figure.

Although the aim of our calculations was merely to obtain results at $\ell\le4000$ that are accurate enough to allow reliable comparison against observational data, it is also interesting to note what expectations we have for $\ell>4000$. As discussed in the previous section, the $\ell\rightarrow\infty$ limit of the \Eq{patchSum} sum suggests the contributions vary as $\ell^{-p}$, where $p=1.5$, $0.9$ and $0.7$ for the scalar, vector and tensor modes respectively. However we caution against large extrapolations using these power-laws since our simulations do not have the dynamical range required to yield confident knowledge of the UETC power-laws at high $k\tau$, particularly in the vector and tensor cases, and that uncertainty feeds through to the present power-laws (see \Fig{ETCextrap}). 


\subsubsection{Comparison with results from Nambu-Goto strings}
\label{sec:compNG}

Firstly our results obtained for small angular scales are in broad agreement with analytical expectations from the GKS effect using a Gaussian model of a Nambu-Goto string network \cite{Hindmarsh:1993pu}, which predicts a $1/\ell$ dependence in the small-scale limit. That model would have yielded ETC functions behaving as $1/k\tau$ at small scales, which is not precisely as seen in our simulations.

Further, this approximate dependence has also been confirmed in work based upon high resolution Nambu-Goto simulations by Fraisse et al. (2008) \cite{Fraisse:2007nu}. While the Nambu-Goto simulation result stems from a method that does not include recombination and hence only includes the effects of strings after last-scattering, their results reveal a $1/\ell^{0.9}$ variation in the power spectrum over the range $400\lesssim\ell\lesssim10^4$. Their calculations are hence most comparable\footnote{Note that in contrast with Fraisse et al., even these results still include the effects of perturbations induced in the matter by the string sources and do not make use of the flat-sky approximation, hence they are still not including exactly the same CMB physics.} to our results in \Fig{CMBpLSo}, when we artificially zeroed our string sources until decoupling, which also indicate a power-law dependence for $\ell\gtrsim400$. Importantly our full results demonstrate for the first time at which angular scales this $\approx1/\ell$ dependence is valid, namely $\ell\gtrsim3000$ --- scales finer than a few arc minutes.

At angular scales near the peak of the spectrum ($\ell\sim400$) there are no recently published results from Nambu-Goto simulations against which we can compare our results, with Contaldi et al. (1999) \cite{Contaldi:1998mx} providing the only example of such work. However, this employed Minkowski space-time simulations plus now out-dated cosmological parameters and therefore a detailed comparison is not appropriate. The basic form of our spectrum is in good agreement with their results, although their power spectrum shows a shift to higher $\ell$ compared to those presented here. Further, their results suggest that Nambu-Goto networks yield CMB predictions with a larger overall normalization. This is also seen by comparing to power spectra calculated from FRW Nambu-Goto simulations and modern cosmological parameters, but which are valid only for discrete multipole ranges (and miss the $\ell\sim400$ peak) \cite{Landriau:2003xf, Landriau:2010cb}. Nambu-Goto strings would require $G\mu\approx0.7\times10^{-6}$ in order to fit observations at $\ell<10$ \cite{Landriau:2003xf}, while we would require $G\mu=1.8\times10^{-6}$ to match the WMAP5 result at $\ell=10$. This is not unexpected given the simulation results and, as shown in \Table{stringXi}, Nambu-Goto calculations yield higher string densities than field theoretic ones, raising the power spectrum normalization and shifting power to smaller scales. We believe this density difference is because field theory simulations provide a decay channel, namely radiative decay, which is not included the Nambu-Goto codes. Despite this, the Fraisse et al. result for $\ell(\ell+1)\mathcal{C}_{\ell}/(G\mu)^2$ at $\ell=4000$ is $\approx25$, which compares favourably to our result of $20$ at this multipole. This may be due to ambiguities associated with their non-inclusion of recombination rather than a real agreement in the amplitude of the power-law.

As discussed in the introduction, the USM is a computationally cheap means of estimating the CMB power spectrum, and this has also been used to study the string contribution at sub-WMAP angular scales with the USM parameters set to approximate a Nambu-Goto network. The published work \cite{Pogosian:2008am} highlighted a $1/\ell^2$ dependence near $\ell \sim 2000$, but a more recent look at greater $\ell$ prompted by our results does indeed reveal an approximately $1/\ell$ variation in the USM
power spectrum above $\ell \sim 3000$ \cite{Pogosian:2008amV2}.


\subsubsection{Comparison with the contribution from inflation}

\begin{figure}
\resizebox{\columnwidth}{!}{\includegraphics{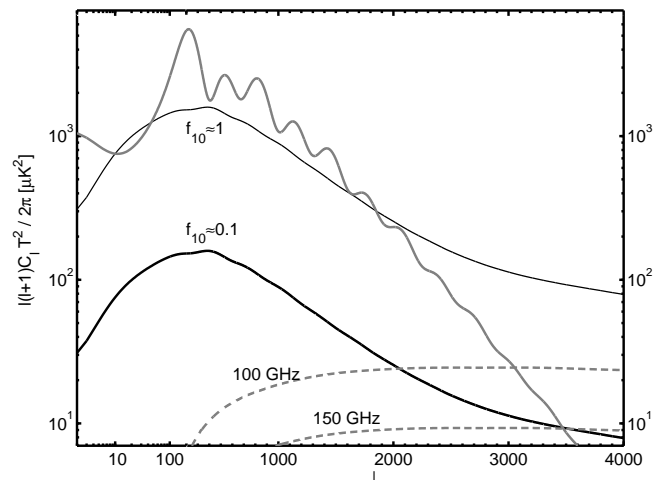}}
\caption{\label{fig:CMBinflation}A comparison of three contributions to the CMB power spectrum: the inflationary scalar contribution (grey, solid), the string contribution (black) and the SZ effect (grey, dashed). The inflationary scalar mode and $\fd\approx1$ string contribution are both normalized to match the WMAP data \cite{Nolta:2008ih} at $\ell=10$, while the $\fd\approx0.1$ string contribution is set to yield 10\% of the observed power at this multipole (the approximation signs are because $\fd$ is defined relative to the total theoretical power spectrum, not the observations). Two normalizations are shown for the SZ contribution, one for each of the QUAD bands: $100$ and $150\unit{GHz}$.}
\end{figure}

It is of course useful to compare the cosmic string power spectrum with that from inflation, to which it should be added\footnote{We calculate the inflationary contribution using CAMB \cite{CAMB=Lewis:1999bs}, with the same parameters as used in the string calculations and also $\ns=1.0$, while assuming negligible contributions from primordial gravity waves.}. This comparison is eased if we give the string contribution an artificially high normalization such that it is equal to the inflationary contribution at $\ell=10$ and then set both to be equal to the value observed by WMAP at that multipole, as in \Fig{CMBinflation}. It can be seen that while the string contribution approximately tracks the troughs in the inflationary contribution for $\ell<1500$, the exponential suppression of the inflationary contribution at high-$\ell$ means that the string component grows in relative size, and dominates for $\ell\gtrsim2000$. Switching to a more realistic string contribution, with one-tenth of the previous normalization, this domination is delayed until $\ell\approx3500$, but importantly the fraction of the total theoretical spectrum due to strings $f_\ell$ increases from $f_{1500}\approx0.1$ to $f_{3500}=0.5$ and therefore accurate data at $\ell\gtrsim2000$ should in principle be highly sensitive to cosmic strings.

However, these are the same angular scales for which the Sunyaev-Zel'dovich (SZ) effect \cite{Sunyaev:1970eu} begins to make a significant contribution to the temperature power spectrum at certain observational frequencies. This contribution results from the distortion of the black-body spectral profile as the CMB photons pass through galaxy clusters and Compton-scatter off hot electrons. In \Fig{CMBinflation} we plot the predictions made by Komatsu and Seljak \cite{Komatsu:2002wc}, normalized\footnote{The plotted SZ spectrum was calculated for $\Obhh=0.023$ and $\sigma_8=0.8$ and scales in normalization as roughly $(\Ob h)^2\sigma_{8}^7$ \cite{Komatsu:2002wc}.} for the two frequency bands observed by the QUAD project \cite{QUAD=Friedman:2009dt}, namely $100$ and $150 \unit{GHz}$. It can be seen that the string contribution with $\fd\approx0.1$ is likely to be shrouded, even at high $\ell$, when measurements are made at low frequencies, but in observational bands near $220\unit{GHz}$, the SZ effect is suppressed. Unfortunately higher frequencies have a greater contribution from unresolved point sources \cite{SPT=Lueker:2009rx} and in practice observations at a number of frequencies will be required in order to understand the frequency dependent contributions and remove them to yield a high sensitivity to the frequency-independent cosmic string component from temperature power spectrum measurements alone. 


\subsection{Polarization Power Spectra}

\begin{figure}
\resizebox{\columnwidth}{!}{\includegraphics{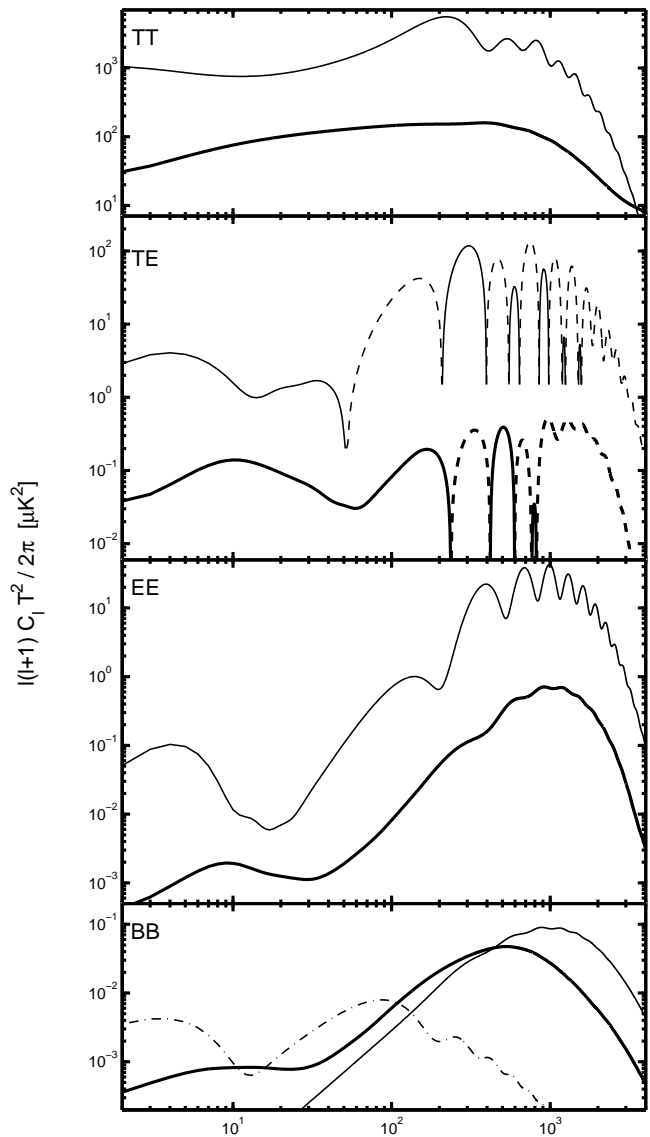}}
\caption{\label{fig:CMBteb} A comparison between the string contribution (thick) to the CMB temperature and polarization power spectra with the adiabatic scalar contribution from inflation (thin). The normalization of the string component is set at $\fd\approx0.1$, while the inflationary scalar contribution uses the same settings as in \Fig{CMBinflation}. For the BB spectrum we also plot the possible contribution from the inflationary tensor mode (dot-dashed), normalized at its $95\%$ upper limit \cite{WMAP7-C:Komatsu:2010fb}.}
\end{figure}

CMB polarization anisotropies differ from those for temperature in that they are created almost exclusively at recombination. Since the polarization is caused by Thompson scattering in the presence of a quadrupole anisotropy, the universe must be ionized in order to create it, but with Thompson scattering still weak enough for it not to have suppressed the quadrupole too heavily. Polarization is hence created only very close to recombination, and also weakly during the recent re-ionized epoch. 
As a result, the polarization signal comes primarily from horizon-scale UETC power impacting at times close to recombination. The contributions from high $k\tau$ near last scattering are outside our window of interest $\ell\le4000$ and are suppressed by Silk-damping. Further, the contributions from high $k\tau$ at recent times are negligible because there is little UETC power at such scales and the re-ionization contribution is secondary in importance. Our present results, shown in \Fig{CMBteb}, therefore add little new information for CMB polarization from strings beyond our previous results \cite{Bevis:2007qz, Urrestilla:2007sf}.

However, for completion, it should be noted that while the string contribution is sub-dominant for all scales in the E-mode polarization spectrum (EE) and the cross correlation of the E-mode with temperature spectrum (TE), it may dominate the B-mode polarization spectrum (BB) for scales $\ell\lesssim400$. This is possible because the inflationary scalar mode contributes to BB only via weak gravitational lensing, which converts EE power into BB. Furthermore, our results indicate that the string contribution peaks at significantly lower $\ell$ than the weak lensing contribution, which should readily prevent confusion between the two, while the weak-lensing contaminant can also be partially removed \cite{Seljak:2006hi}. Note that we find a BB peak from strings that is at larger scales than is seen in USM results when that model is set to approximate Nambu-Goto simulations \cite{Pogosian:2007gi}, as expected given our lower string densities, and that the USM results suggest more confusion with the weak-lensing signal. It is also possible that the inflationary tensor mode may make a sizable contribution to the BB spectrum at very large angular scales, as shown in \Fig{CMBteb}, but its normalization depends on the details of the inflationary model and is poorly constrained by current data. A wealth of B-mode data will soon be available and promises to be highly sensitive to cosmic strings \cite{Seljak:2006hi, Bevis:2007qz, Pogosian:2007gi}.


\section{Conclusions}
\label{sec:conclusions}

We have presented the first calculation of the cosmic string CMB temperature power spectrum contribution that is accurate over the multipole range $\ell=2\rightarrow4000$, a range that encloses the scales probed by the Planck satellite ($\ell=2\rightarrow2500$) and additionally scales at which sub-orbital data are becoming increasingly accurate. Further, we show that at $\ell\sim3000$ there is a knee in the power spectrum result due to the exponential decay of the early time contribution at small-scales, which then reveals the post-recombination component. This late-time contribution varies as roughly $1/\ell$, which is the basic expectation for the GKS effect \cite{Hindmarsh:1993pu}. Our results yield values for the power law exponent for each of the scalar, vector and tensor contributions to the overall high-$\ell$ behaviour and can in principle be used to estimate the temperature contribution from strings for $\ell>4000$. However this extrapolation must be performed with caution since our simulations do not have the dynamic range to yield these power laws with great confidence. 

Our results indicate that the size of the cosmic string signal in the temperature power spectrum, relative the adiabatic inflationary component, is roughly 10 times greater at $\ell\approx3500$ than it is at $\ell\approx10$ or $\ell\approx400-1500$ and therefore small scales are particularly important for cosmic strings. While it is true that other contributions also become significant at small scales, for example the Sunyaev-Zel'dovich effect, such effects are frequency-dependent and hence can be identified and subtracted. Further, our results are not limited to temperature anisotropies but include polarization also, which is of great future importance for strings in the case of the B-mode. We present a comparison of our temperature and polarization results to the latest data in a separate, shorter article \cite{Bevis:2010data}.

Our simulations also yield greater knowledge of the scaling properties of Abelian Higgs string networks, with improvements in the accuracy of results such as the scaling density and the UETCs. Additionally, we have measured the coherence function for Abelian Higgs strings for the first time, the form of which is important for calculations of this kind.

That our results stem from the Abelian Higgs model means, of course, that they are not necessarily accurate for cosmic superstrings (see eg. \cite{Copeland:2009ga} for a review). These superstrings may have intercommutation probabilities significantly lower than that for gauge strings and additionally may form Y-shaped junctions, which do not form in the Abelian Higgs model for the parameters chosen here. While small-scale structure on the strings has been shown to lessen the impact of the intercommutation probability \cite{Avgoustidis:2005nv}, the effect of Y-junctions on the network properties is highly uncertain. Both effects are likely to increase the string density and decrease the inter-string distance, resulting in a shift of the peak in the CMB signal to greater $\ell$ values. This may therefore enable superstrings to be distinguished from conventional cosmic strings should a string component in the CMB be detected in future data.

Finally, we note that the power spectrum is not a complete statistical description of the anisotropies that would be seeded by cosmic strings, since there would be a significant non-Gaussian character created by the GKS effect. Higher order moments such as the bispectrum and trispectrum have been calculated \cite{Hindmarsh:2009qk, Hindmarsh:2009es, Regan:2009hv}, in addition to realizations of CMB maps \cite{Ringeval:2005kr, Landriau:2010cb}. While these calculations are challenging and either do not include recombination or are valid only for very limited multipole ranges, non-Gaussianity is an exciting channel for future cosmic string constraint or detection.


\begin{acknowledgments}
We acknowledge financial support from STFC (N.B.), the Swiss National Science Foundation (M.K.), the Spanish Consolider-Ingenio 2010 Programme CPAN (CSD2007-00042), and the Spanish Ministry of Science and Innovation FPA2009-10612 (J.U.). Numerical calculations were performed using the UK National Cosmology Supercomputer (COSMOS, supported by SGI/Intel, HEFCE, and STFC), the Imperial College HPC facilities, and the University of Sussex HPC Archimedes cluster. We thank Arttu Rajantie and Levon Pogosian for useful discussions.
\end{acknowledgments}

\bibliography{references}

\begin{thebibliography}{79}
\expandafter\ifx\csname natexlab\endcsname\relax\def\natexlab#1{#1}\fi
\expandafter\ifx\csname bibnamefont\endcsname\relax
  \def\bibnamefont#1{#1}\fi
\expandafter\ifx\csname bibfnamefont\endcsname\relax
  \def\bibfnamefont#1{#1}\fi
\expandafter\ifx\csname citenamefont\endcsname\relax
  \def\citenamefont#1{#1}\fi
\expandafter\ifx\csname url\endcsname\relax
  \def\url#1{\texttt{#1}}\fi
\expandafter\ifx\csname urlprefix\endcsname\relax\def\urlprefix{URL }\fi
\providecommand{\bibinfo}[2]{#2}
\providecommand{\eprint}[2][]{\url{#2}}

\bibitem[{\citenamefont{Nolta et~al.}(2009)}]{Nolta:2008ih}
\bibinfo{author}{\bibfnamefont{M.~R.} \bibnamefont{Nolta}} \bibnamefont{et~al.}
  (\bibinfo{collaboration}{WMAP}), \bibinfo{journal}{Astrophys. J. Suppl.}
  \textbf{\bibinfo{volume}{180}}, \bibinfo{pages}{296} (\bibinfo{year}{2009}),
  \eprint{0803.0593}.

\bibitem[{\citenamefont{Percival et~al.}(2009)}]{Percival:2009xn}
\bibinfo{author}{\bibfnamefont{W.~J.} \bibnamefont{Percival}}
  \bibnamefont{et~al.} (\bibinfo{year}{2009}), \eprint{0907.1660}.

\bibitem[{\citenamefont{Bevis et~al.}(2008)\citenamefont{Bevis, Hindmarsh,
  Kunz, and Urrestilla}}]{Bevis:2007gh}
\bibinfo{author}{\bibfnamefont{N.}~\bibnamefont{Bevis}},
  \bibinfo{author}{\bibfnamefont{M.}~\bibnamefont{Hindmarsh}},
  \bibinfo{author}{\bibfnamefont{M.}~\bibnamefont{Kunz}}, \bibnamefont{and}
  \bibinfo{author}{\bibfnamefont{J.}~\bibnamefont{Urrestilla}},
  \bibinfo{journal}{Phys. Rev. Lett.} \textbf{\bibinfo{volume}{100}},
  \bibinfo{pages}{021301} (\bibinfo{year}{2008}), \eprint{astro-ph/0702223}.

\bibitem[{\citenamefont{Battye et~al.}(2006)\citenamefont{Battye, Garbrecht,
  and Moss}}]{Battye:2006pk}
\bibinfo{author}{\bibfnamefont{R.~A.} \bibnamefont{Battye}},
  \bibinfo{author}{\bibfnamefont{B.}~\bibnamefont{Garbrecht}},
  \bibnamefont{and} \bibinfo{author}{\bibfnamefont{A.}~\bibnamefont{Moss}},
  \bibinfo{journal}{JCAP} \textbf{\bibinfo{volume}{0609}}, \bibinfo{pages}{007}
  (\bibinfo{year}{2006}), \eprint{astro-ph/0607339}.

\bibitem[{\citenamefont{Fraisse}(2007)}]{Fraisse:2006xc}
\bibinfo{author}{\bibfnamefont{A.~A.} \bibnamefont{Fraisse}},
  \bibinfo{journal}{JCAP} \textbf{\bibinfo{volume}{0703}}, \bibinfo{pages}{008}
  (\bibinfo{year}{2007}), \eprint{astro-ph/0603589}.

\bibitem[{\citenamefont{Wyman et~al.}(2005)\citenamefont{Wyman, Pogosian, and
  Wasserman}}]{Wyman:2005tu}
\bibinfo{author}{\bibfnamefont{M.}~\bibnamefont{Wyman}},
  \bibinfo{author}{\bibfnamefont{L.}~\bibnamefont{Pogosian}}, \bibnamefont{and}
  \bibinfo{author}{\bibfnamefont{I.}~\bibnamefont{Wasserman}},
  \bibinfo{journal}{Phys. Rev.} \textbf{\bibinfo{volume}{D72}},
  \bibinfo{pages}{023513} (\bibinfo{year}{2005}), \eprint{astro-ph/0503364}.

\bibitem[{\citenamefont{Vilenkin and Shellard}(1994)}]{Vilenkin:1994book}
\bibinfo{author}{\bibfnamefont{A.}~\bibnamefont{Vilenkin}} \bibnamefont{and}
  \bibinfo{author}{\bibfnamefont{E.~P.~S.} \bibnamefont{Shellard}},
  \emph{\bibinfo{title}{Cosmic Strings and Other Topological Defects}}
  (\bibinfo{publisher}{Cambridge University Press, Cambridge, U.K.},
  \bibinfo{year}{1994}).

\bibitem[{\citenamefont{Hindmarsh and Kibble}(1995)}]{Hindmarsh:1994re}
\bibinfo{author}{\bibfnamefont{M.~B.} \bibnamefont{Hindmarsh}}
  \bibnamefont{and} \bibinfo{author}{\bibfnamefont{T.~W.~B.}
  \bibnamefont{Kibble}}, \bibinfo{journal}{Rept. Prog. Phys.}
  \textbf{\bibinfo{volume}{58}}, \bibinfo{pages}{477} (\bibinfo{year}{1995}),
  \eprint{hep-ph/9411342}.

\bibitem[{\citenamefont{Sakellariadou}(2007)}]{Sakellariadou:2006qs}
\bibinfo{author}{\bibfnamefont{M.}~\bibnamefont{Sakellariadou}},
  \bibinfo{journal}{Lect. Notes Phys.} \textbf{\bibinfo{volume}{718}},
  \bibinfo{pages}{247} (\bibinfo{year}{2007}), \eprint{hep-th/0602276}.

\bibitem[{\citenamefont{Copeland and Kibble}(2009)}]{Copeland:2009ga}
\bibinfo{author}{\bibfnamefont{E.~J.} \bibnamefont{Copeland}} \bibnamefont{and}
  \bibinfo{author}{\bibfnamefont{T.~W.~B.} \bibnamefont{Kibble}}
  (\bibinfo{year}{2009}), \eprint{0911.1345}.

\bibitem[{\citenamefont{Sarangi and Tye}(2002)}]{Sarangi:2002yt}
\bibinfo{author}{\bibfnamefont{S.}~\bibnamefont{Sarangi}} \bibnamefont{and}
  \bibinfo{author}{\bibfnamefont{S.~H.~H.} \bibnamefont{Tye}},
  \bibinfo{journal}{Phys. Lett.} \textbf{\bibinfo{volume}{B536}},
  \bibinfo{pages}{185} (\bibinfo{year}{2002}), \eprint{hep-th/0204074}.

\bibitem[{\citenamefont{Jones et~al.}(2003)\citenamefont{Jones, Stoica, and
  Tye}}]{Jones:2003da}
\bibinfo{author}{\bibfnamefont{N.~T.} \bibnamefont{Jones}},
  \bibinfo{author}{\bibfnamefont{H.}~\bibnamefont{Stoica}}, \bibnamefont{and}
  \bibinfo{author}{\bibfnamefont{S.~H.~H.} \bibnamefont{Tye}},
  \bibinfo{journal}{Phys. Lett.} \textbf{\bibinfo{volume}{B563}},
  \bibinfo{pages}{6} (\bibinfo{year}{2003}), \eprint{hep-th/0303269}.

\bibitem[{\citenamefont{Jeannerot et~al.}(2003)\citenamefont{Jeannerot, Rocher,
  and Sakellariadou}}]{Jeannerot:2003qv}
\bibinfo{author}{\bibfnamefont{R.}~\bibnamefont{Jeannerot}},
  \bibinfo{author}{\bibfnamefont{J.}~\bibnamefont{Rocher}}, \bibnamefont{and}
  \bibinfo{author}{\bibfnamefont{M.}~\bibnamefont{Sakellariadou}},
  \bibinfo{journal}{Phys. Rev.} \textbf{\bibinfo{volume}{D68}},
  \bibinfo{pages}{103514} (\bibinfo{year}{2003}), \eprint{hep-ph/0308134}.

\bibitem[{\citenamefont{Vachaspati and Achucarro}(1991)}]{Vachaspati:1991dz}
\bibinfo{author}{\bibfnamefont{T.}~\bibnamefont{Vachaspati}} \bibnamefont{and}
  \bibinfo{author}{\bibfnamefont{A.}~\bibnamefont{Achucarro}},
  \bibinfo{journal}{Phys. Rev.} \textbf{\bibinfo{volume}{D44}},
  \bibinfo{pages}{3067} (\bibinfo{year}{1991}).

\bibitem[{\citenamefont{Hindmarsh}(1992)}]{Hindmarsh:1991jq}
\bibinfo{author}{\bibfnamefont{M.}~\bibnamefont{Hindmarsh}},
  \bibinfo{journal}{Phys. Rev. Lett.} \textbf{\bibinfo{volume}{68}},
  \bibinfo{pages}{1263} (\bibinfo{year}{1992}).

\bibitem[{\citenamefont{Ach\'ucarro and Vachaspati}(2000)}]{Achucarro:1999it}
\bibinfo{author}{\bibfnamefont{A.}~\bibnamefont{Ach\'ucarro}} \bibnamefont{and}
  \bibinfo{author}{\bibfnamefont{T.}~\bibnamefont{Vachaspati}},
  \bibinfo{journal}{Phys. Rept.} \textbf{\bibinfo{volume}{327}},
  \bibinfo{pages}{347} (\bibinfo{year}{2000}), \eprint{hep-ph/9904229}.

\bibitem[{\citenamefont{Ach\'ucarro et~al.}(2007)\citenamefont{Ach\'ucarro,
  Salmi, and Urrestilla}}]{Achucarro:2007sp}
\bibinfo{author}{\bibfnamefont{A.}~\bibnamefont{Ach\'ucarro}},
  \bibinfo{author}{\bibfnamefont{P.}~\bibnamefont{Salmi}}, \bibnamefont{and}
  \bibinfo{author}{\bibfnamefont{J.}~\bibnamefont{Urrestilla}},
  \bibinfo{journal}{Phys. Rev.} \textbf{\bibinfo{volume}{D75}},
  \bibinfo{pages}{121703} (\bibinfo{year}{2007}), \eprint{astro-ph/0512487}.

\bibitem[{\citenamefont{Pen et~al.}(1997)\citenamefont{Pen, Seljak, and
  Turok}}]{Pen:1997ae}
\bibinfo{author}{\bibfnamefont{U.-L.} \bibnamefont{Pen}},
  \bibinfo{author}{\bibfnamefont{U.}~\bibnamefont{Seljak}}, \bibnamefont{and}
  \bibinfo{author}{\bibfnamefont{N.}~\bibnamefont{Turok}},
  \bibinfo{journal}{Phys. Rev. Lett.} \textbf{\bibinfo{volume}{79}},
  \bibinfo{pages}{1611} (\bibinfo{year}{1997}), \eprint{astro-ph/9704165}.

\bibitem[{\citenamefont{Durrer et~al.}(1999)\citenamefont{Durrer, Kunz, and
  Melchiorri}}]{Durrer:1998rw}
\bibinfo{author}{\bibfnamefont{R.}~\bibnamefont{Durrer}},
  \bibinfo{author}{\bibfnamefont{M.}~\bibnamefont{Kunz}}, \bibnamefont{and}
  \bibinfo{author}{\bibfnamefont{A.}~\bibnamefont{Melchiorri}},
  \bibinfo{journal}{Phys. Rev.} \textbf{\bibinfo{volume}{D59}},
  \bibinfo{pages}{123005} (\bibinfo{year}{1999}), \eprint{astro-ph/9811174}.

\bibitem[{\citenamefont{Bevis et~al.}(2004)\citenamefont{Bevis, Hindmarsh, and
  Kunz}}]{Bevis:2004wk}
\bibinfo{author}{\bibfnamefont{N.}~\bibnamefont{Bevis}},
  \bibinfo{author}{\bibfnamefont{M.}~\bibnamefont{Hindmarsh}},
  \bibnamefont{and} \bibinfo{author}{\bibfnamefont{M.}~\bibnamefont{Kunz}},
  \bibinfo{journal}{Phys. Rev.} \textbf{\bibinfo{volume}{D70}},
  \bibinfo{pages}{043508} (\bibinfo{year}{2004}), \eprint{astro-ph/0403029}.

\bibitem[{\citenamefont{Cruz et~al.}(2007)\citenamefont{Cruz, Turok, Vielva,
  Martinez-Gonzalez, and Hobson}}]{Cruz:2007pe}
\bibinfo{author}{\bibfnamefont{M.}~\bibnamefont{Cruz}},
  \bibinfo{author}{\bibfnamefont{N.}~\bibnamefont{Turok}},
  \bibinfo{author}{\bibfnamefont{P.}~\bibnamefont{Vielva}},
  \bibinfo{author}{\bibfnamefont{E.}~\bibnamefont{Martinez-Gonzalez}},
  \bibnamefont{and} \bibinfo{author}{\bibfnamefont{M.}~\bibnamefont{Hobson}},
  \bibinfo{journal}{Science} \textbf{\bibinfo{volume}{318}},
  \bibinfo{pages}{1612} (\bibinfo{year}{2007}), \eprint{0710.5737}.

\bibitem[{\citenamefont{Urrestilla et~al.}(2008)\citenamefont{Urrestilla,
  Bevis, Hindmarsh, Kunz, and Liddle}}]{Urrestilla:2007sf}
\bibinfo{author}{\bibfnamefont{J.}~\bibnamefont{Urrestilla}},
  \bibinfo{author}{\bibfnamefont{N.}~\bibnamefont{Bevis}},
  \bibinfo{author}{\bibfnamefont{M.}~\bibnamefont{Hindmarsh}},
  \bibinfo{author}{\bibfnamefont{M.}~\bibnamefont{Kunz}}, \bibnamefont{and}
  \bibinfo{author}{\bibfnamefont{A.~R.} \bibnamefont{Liddle}},
  \bibinfo{journal}{JCAP} \textbf{\bibinfo{volume}{0807}}, \bibinfo{pages}{010}
  (\bibinfo{year}{2008}), \eprint{arXiv:0711.1842 [astro-ph]}.

\bibitem[{\citenamefont{Bevis et~al.}(2007{\natexlab{a}})\citenamefont{Bevis,
  Hindmarsh, Kunz, and Urrestilla}}]{Bevis:2006mj}
\bibinfo{author}{\bibfnamefont{N.}~\bibnamefont{Bevis}},
  \bibinfo{author}{\bibfnamefont{M.}~\bibnamefont{Hindmarsh}},
  \bibinfo{author}{\bibfnamefont{M.}~\bibnamefont{Kunz}}, \bibnamefont{and}
  \bibinfo{author}{\bibfnamefont{J.}~\bibnamefont{Urrestilla}},
  \bibinfo{journal}{Phys. Rev.} \textbf{\bibinfo{volume}{D75}},
  \bibinfo{pages}{065015} (\bibinfo{year}{2007}{\natexlab{a}}),
  \eprint{astro-ph/0605018}.

\bibitem[{\citenamefont{Bevis et~al.}(2007{\natexlab{b}})\citenamefont{Bevis,
  Hindmarsh, Kunz, and Urrestilla}}]{Bevis:2007qz}
\bibinfo{author}{\bibfnamefont{N.}~\bibnamefont{Bevis}},
  \bibinfo{author}{\bibfnamefont{M.}~\bibnamefont{Hindmarsh}},
  \bibinfo{author}{\bibfnamefont{M.}~\bibnamefont{Kunz}}, \bibnamefont{and}
  \bibinfo{author}{\bibfnamefont{J.}~\bibnamefont{Urrestilla}},
  \bibinfo{journal}{Phys. Rev.} \textbf{\bibinfo{volume}{D76}},
  \bibinfo{pages}{043005} (\bibinfo{year}{2007}{\natexlab{b}}),
  \eprint{arXiv:0704.3800 [astro-ph]}.

\bibitem[{\citenamefont{Hinshaw and others
  (WMAP)}(2006)}]{WMAP3-T=Hinshaw:2006ia}
\bibinfo{author}{\bibfnamefont{G.}~\bibnamefont{Hinshaw}} \bibnamefont{and}
  \bibinfo{author}{\bibnamefont{others (WMAP)}} (\bibinfo{year}{2006}),
  \eprint{astro-ph/0603451}.

\bibitem[{\citenamefont{Spergel and others
  (WMAP)}(2006)}]{WMAP3-C=Spergel:2006hy}
\bibinfo{author}{\bibfnamefont{D.~N.} \bibnamefont{Spergel}} \bibnamefont{and}
  \bibinfo{author}{\bibnamefont{others (WMAP)}} (\bibinfo{year}{2006}),
  \eprint{astro-ph/0603449}.

\bibitem[{\citenamefont{Sievers et~al.}(2009)}]{CBI=Sievers:2009ah}
\bibinfo{author}{\bibfnamefont{J.~L.} \bibnamefont{Sievers}}
  \bibnamefont{et~al.} (\bibinfo{collaboration}{CBI}) (\bibinfo{year}{2009}),
  \eprint{arXiv:0901.4540}.

\bibitem[{\citenamefont{Reichardt et~al.}(2009)}]{ACBAR=Reichardt:2008ay}
\bibinfo{author}{\bibfnamefont{C.~L.} \bibnamefont{Reichardt}}
  \bibnamefont{et~al.} (\bibinfo{collaboration}{ACBAR}),
  \bibinfo{journal}{Astrophys. J.} \textbf{\bibinfo{volume}{694}},
  \bibinfo{pages}{1200} (\bibinfo{year}{2009}), \eprint{0801.1491}.

\bibitem[{\citenamefont{Friedman et~al.}(2009)}]{QUAD=Friedman:2009dt}
\bibinfo{author}{\bibfnamefont{R.~B.} \bibnamefont{Friedman}}
  \bibnamefont{et~al.} (\bibinfo{collaboration}{QUaD}),
  \bibinfo{journal}{Astrophys. J.} \textbf{\bibinfo{volume}{700}},
  \bibinfo{pages}{L187} (\bibinfo{year}{2009}), \eprint{0901.4334}.

\bibitem[{\citenamefont{Lueker et~al.}(2009)}]{SPT=Lueker:2009rx}
\bibinfo{author}{\bibfnamefont{M.}~\bibnamefont{Lueker}} \bibnamefont{et~al.}
  (\bibinfo{year}{2009}), \eprint{0912.4317}.

\bibitem[{\citenamefont{Fowler et~al.}(2010)}]{ACT=Fowler:2010cy}
\bibinfo{author}{\bibfnamefont{J.}~\bibnamefont{Fowler}} \bibnamefont{et~al.}
  (\bibinfo{collaboration}{The ACT}) (\bibinfo{year}{2010}),
  \eprint{1001.2934}.

\bibitem[{Pla()}]{Planck}
\bibinfo{note}{Planck satelite, ESA},
  \urlprefix\url{http://www.rssd.esa.int/index.php?project=Planck}.

\bibitem[{\citenamefont{Hindmarsh}(1994)}]{Hindmarsh:1993pu}
\bibinfo{author}{\bibfnamefont{M.}~\bibnamefont{Hindmarsh}},
  \bibinfo{journal}{Astrophys. J.} \textbf{\bibinfo{volume}{431}},
  \bibinfo{pages}{534} (\bibinfo{year}{1994}), \eprint{astro-ph/9307040}.

\bibitem[{\citenamefont{Kibble}(1985)}]{Kibble:1984hp}
\bibinfo{author}{\bibfnamefont{T.~W.~B.} \bibnamefont{Kibble}},
  \bibinfo{journal}{Nucl. Phys.} \textbf{\bibinfo{volume}{B252}},
  \bibinfo{pages}{227} (\bibinfo{year}{1985}).

\bibitem[{\citenamefont{Kaiser and Stebbins}(1984)}]{Kaiser:1984iv}
\bibinfo{author}{\bibfnamefont{N.}~\bibnamefont{Kaiser}} \bibnamefont{and}
  \bibinfo{author}{\bibfnamefont{A.}~\bibnamefont{Stebbins}},
  \bibinfo{journal}{Nature} \textbf{\bibinfo{volume}{310}},
  \bibinfo{pages}{391} (\bibinfo{year}{1984}).

\bibitem[{\citenamefont{Stebbins}(1988)}]{Stebbins:1987va}
\bibinfo{author}{\bibfnamefont{A.}~\bibnamefont{Stebbins}},
  \bibinfo{journal}{Astrophys. J.} \textbf{\bibinfo{volume}{327}},
  \bibinfo{pages}{584} (\bibinfo{year}{1988}).

\bibitem[{\citenamefont{Gott}(1985)}]{Gott:1984ef}
\bibinfo{author}{\bibfnamefont{J.~R.} \bibnamefont{Gott}, \bibfnamefont{III}},
  \bibinfo{journal}{Astrophys. J.} \textbf{\bibinfo{volume}{288}},
  \bibinfo{pages}{422} (\bibinfo{year}{1985}).

\bibitem[{\citenamefont{Landriau and Shellard}(2004)}]{Landriau:2003xf}
\bibinfo{author}{\bibfnamefont{M.}~\bibnamefont{Landriau}} \bibnamefont{and}
  \bibinfo{author}{\bibfnamefont{E.~P.~S.} \bibnamefont{Shellard}},
  \bibinfo{journal}{Phys. Rev.} \textbf{\bibinfo{volume}{D69}},
  \bibinfo{pages}{023003} (\bibinfo{year}{2004}), \eprint{astro-ph/0302166}.

\bibitem[{\citenamefont{Fraisse et~al.}(2008)\citenamefont{Fraisse, Ringeval,
  Spergel, and Bouchet}}]{Fraisse:2007nu}
\bibinfo{author}{\bibfnamefont{A.~A.} \bibnamefont{Fraisse}},
  \bibinfo{author}{\bibfnamefont{C.}~\bibnamefont{Ringeval}},
  \bibinfo{author}{\bibfnamefont{D.~N.} \bibnamefont{Spergel}},
  \bibnamefont{and} \bibinfo{author}{\bibfnamefont{F.~R.}
  \bibnamefont{Bouchet}}, \bibinfo{journal}{Phys. Rev.}
  \textbf{\bibinfo{volume}{D78}}, \bibinfo{pages}{043535}
  (\bibinfo{year}{2008}), \eprint{0708.1162}.

\bibitem[{\citenamefont{Contaldi et~al.}(1999)\citenamefont{Contaldi,
  Hindmarsh, and Magueijo}}]{Contaldi:1998mx}
\bibinfo{author}{\bibfnamefont{C.}~\bibnamefont{Contaldi}},
  \bibinfo{author}{\bibfnamefont{M.}~\bibnamefont{Hindmarsh}},
  \bibnamefont{and} \bibinfo{author}{\bibfnamefont{J.}~\bibnamefont{Magueijo}},
  \bibinfo{journal}{Phys. Rev. Lett.} \textbf{\bibinfo{volume}{82}},
  \bibinfo{pages}{679} (\bibinfo{year}{1999}), \eprint{astro-ph/9808201}.

\bibitem[{\citenamefont{Albrecht et~al.}(1997)\citenamefont{Albrecht, Battye,
  and Robinson}}]{Albrecht:1997nt}
\bibinfo{author}{\bibfnamefont{A.}~\bibnamefont{Albrecht}},
  \bibinfo{author}{\bibfnamefont{R.~A.} \bibnamefont{Battye}},
  \bibnamefont{and} \bibinfo{author}{\bibfnamefont{J.}~\bibnamefont{Robinson}},
  \bibinfo{journal}{Phys. Rev. Lett.} \textbf{\bibinfo{volume}{79}},
  \bibinfo{pages}{4736} (\bibinfo{year}{1997}), \eprint{astro-ph/9707129}.

\bibitem[{\citenamefont{Pogosian and Vachaspati}(1999)}]{Pogosian:1999np}
\bibinfo{author}{\bibfnamefont{L.}~\bibnamefont{Pogosian}} \bibnamefont{and}
  \bibinfo{author}{\bibfnamefont{T.}~\bibnamefont{Vachaspati}},
  \bibinfo{journal}{Phys. Rev.} \textbf{\bibinfo{volume}{D60}},
  \bibinfo{pages}{083504} (\bibinfo{year}{1999}), \eprint{astro-ph/9903361}.

\bibitem[{\citenamefont{Pogosian et~al.}(2008)\citenamefont{Pogosian, Tye,
  Wasserman, and Wyman}}]{Pogosian:2008am}
\bibinfo{author}{\bibfnamefont{L.}~\bibnamefont{Pogosian}},
  \bibinfo{author}{\bibfnamefont{S.~H.~H.} \bibnamefont{Tye}},
  \bibinfo{author}{\bibfnamefont{I.}~\bibnamefont{Wasserman}},
  \bibnamefont{and} \bibinfo{author}{\bibfnamefont{M.}~\bibnamefont{Wyman}}
  (\bibinfo{year}{2008}).

\bibitem[{\citenamefont{Ringeval et~al.}(2007)\citenamefont{Ringeval,
  Sakellariadou, and Bouchet}}]{Ringeval:2005kr}
\bibinfo{author}{\bibfnamefont{C.}~\bibnamefont{Ringeval}},
  \bibinfo{author}{\bibfnamefont{M.}~\bibnamefont{Sakellariadou}},
  \bibnamefont{and} \bibinfo{author}{\bibfnamefont{F.}~\bibnamefont{Bouchet}},
  \bibinfo{journal}{JCAP} \textbf{\bibinfo{volume}{0702}}, \bibinfo{pages}{023}
  (\bibinfo{year}{2007}), \eprint{astro-ph/0511646}.

\bibitem[{\citenamefont{Olum and Vanchurin}(2007)}]{Olum:2006ix}
\bibinfo{author}{\bibfnamefont{K.~D.} \bibnamefont{Olum}} \bibnamefont{and}
  \bibinfo{author}{\bibfnamefont{V.}~\bibnamefont{Vanchurin}},
  \bibinfo{journal}{Phys. Rev.} \textbf{\bibinfo{volume}{D75}},
  \bibinfo{pages}{063521} (\bibinfo{year}{2007}), \eprint{astro-ph/0610419}.

\bibitem[{\citenamefont{Pogosian and Wyman}(2008)}]{Pogosian:2007gi}
\bibinfo{author}{\bibfnamefont{L.}~\bibnamefont{Pogosian}} \bibnamefont{and}
  \bibinfo{author}{\bibfnamefont{M.}~\bibnamefont{Wyman}},
  \bibinfo{journal}{Phys. Rev.} \textbf{\bibinfo{volume}{D77}},
  \bibinfo{pages}{083509} (\bibinfo{year}{2008}), \eprint{arXiv:0711.0747
  [astro-ph]}.

\bibitem[{\citenamefont{Battye et~al.}(2010)\citenamefont{Battye, Garbrecht,
  and Moss}}]{Battye:2010hg}
\bibinfo{author}{\bibfnamefont{R.}~\bibnamefont{Battye}},
  \bibinfo{author}{\bibfnamefont{B.}~\bibnamefont{Garbrecht}},
  \bibnamefont{and} \bibinfo{author}{\bibfnamefont{A.}~\bibnamefont{Moss}}
  (\bibinfo{year}{2010}), \eprint{1001.0769}.

\bibitem[{\citenamefont{Battye and Moss}(2010)}]{Battye:2010xz}
\bibinfo{author}{\bibfnamefont{R.}~\bibnamefont{Battye}} \bibnamefont{and}
  \bibinfo{author}{\bibfnamefont{A.}~\bibnamefont{Moss}}
  (\bibinfo{year}{2010}), \eprint{1005.0479}.

\bibitem[{\citenamefont{Vincent et~al.}(1998)\citenamefont{Vincent, Antunes,
  and Hindmarsh}}]{Vincent:1997cx}
\bibinfo{author}{\bibfnamefont{G.}~\bibnamefont{Vincent}},
  \bibinfo{author}{\bibfnamefont{N.~D.} \bibnamefont{Antunes}},
  \bibnamefont{and}
  \bibinfo{author}{\bibfnamefont{M.}~\bibnamefont{Hindmarsh}},
  \bibinfo{journal}{Phys. Rev. Lett.} \textbf{\bibinfo{volume}{80}},
  \bibinfo{pages}{2277} (\bibinfo{year}{1998}), \eprint{hep-ph/9708427}.

\bibitem[{\citenamefont{Hindmarsh
  et~al.}(2009{\natexlab{a}})\citenamefont{Hindmarsh, Stuckey, and
  Bevis}}]{Hindmarsh:2008dw}
\bibinfo{author}{\bibfnamefont{M.}~\bibnamefont{Hindmarsh}},
  \bibinfo{author}{\bibfnamefont{S.}~\bibnamefont{Stuckey}}, \bibnamefont{and}
  \bibinfo{author}{\bibfnamefont{N.}~\bibnamefont{Bevis}},
  \bibinfo{journal}{Phys. Rev.} \textbf{\bibinfo{volume}{D79}},
  \bibinfo{pages}{123504} (\bibinfo{year}{2009}{\natexlab{a}}),
  \eprint{0812.1929}.

\bibitem[{\citenamefont{Seljak and Zaldarriaga}(1996)}]{Seljak:1996is}
\bibinfo{author}{\bibfnamefont{U.}~\bibnamefont{Seljak}} \bibnamefont{and}
  \bibinfo{author}{\bibfnamefont{M.}~\bibnamefont{Zaldarriaga}},
  \bibinfo{journal}{Astrophys. J.} \textbf{\bibinfo{volume}{469}},
  \bibinfo{pages}{437} (\bibinfo{year}{1996}), \eprint{astro-ph/9603033}.

\bibitem[{\citenamefont{Martins and Shellard}(2006)}]{Martins:2005es}
\bibinfo{author}{\bibfnamefont{C.~J. A.~P.} \bibnamefont{Martins}}
  \bibnamefont{and} \bibinfo{author}{\bibfnamefont{E.~P.~S.}
  \bibnamefont{Shellard}}, \bibinfo{journal}{Phys. Rev.}
  \textbf{\bibinfo{volume}{D73}}, \bibinfo{pages}{043515}
  (\bibinfo{year}{2006}), \eprint{astro-ph/0511792}.

\bibitem[{\citenamefont{Moore et~al.}(2001)\citenamefont{Moore, Shellard, and
  Martins}}]{Moore:2001px}
\bibinfo{author}{\bibfnamefont{J.~N.} \bibnamefont{Moore}},
  \bibinfo{author}{\bibfnamefont{E.~P.~S.} \bibnamefont{Shellard}},
  \bibnamefont{and} \bibinfo{author}{\bibfnamefont{C.~J. A.~P.}
  \bibnamefont{Martins}}, \bibinfo{journal}{Phys. Rev.}
  \textbf{\bibinfo{volume}{D65}}, \bibinfo{pages}{023503}
  (\bibinfo{year}{2001}), \eprint{hep-ph/0107171}.

\bibitem[{\citenamefont{Turok}(1996)}]{Turok:1996ud}
\bibinfo{author}{\bibfnamefont{N.}~\bibnamefont{Turok}},
  \bibinfo{journal}{Phys. Rev.} \textbf{\bibinfo{volume}{D54}},
  \bibinfo{pages}{R3686} (\bibinfo{year}{1996}), \eprint{astro-ph/9604172}.

\bibitem[{\citenamefont{Doran}(2005)}]{Doran:2003sy}
\bibinfo{author}{\bibfnamefont{M.}~\bibnamefont{Doran}},
  \bibinfo{journal}{JCAP} \textbf{\bibinfo{volume}{0510}}, \bibinfo{pages}{011}
  (\bibinfo{year}{2005}), \eprint{astro-ph/0302138}.

\bibitem[{\citenamefont{Durrer and Kunz}(1998)}]{Durrer:1997ep}
\bibinfo{author}{\bibfnamefont{R.}~\bibnamefont{Durrer}} \bibnamefont{and}
  \bibinfo{author}{\bibfnamefont{M.}~\bibnamefont{Kunz}},
  \bibinfo{journal}{Phys. Rev.} \textbf{\bibinfo{volume}{D57}},
  \bibinfo{pages}{R3199} (\bibinfo{year}{1998}), \eprint{astro-ph/9711133}.

\bibitem[{\citenamefont{Bardeen}(1980)}]{Bardeen:1980kt}
\bibinfo{author}{\bibfnamefont{J.~M.} \bibnamefont{Bardeen}},
  \bibinfo{journal}{Phys. Rev.} \textbf{\bibinfo{volume}{D22}},
  \bibinfo{pages}{1882} (\bibinfo{year}{1980}).

\bibitem[{\citenamefont{Durrer et~al.}(2002)\citenamefont{Durrer, Kunz, and
  Melchiorri}}]{Durrer:2001cg}
\bibinfo{author}{\bibfnamefont{R.}~\bibnamefont{Durrer}},
  \bibinfo{author}{\bibfnamefont{M.}~\bibnamefont{Kunz}}, \bibnamefont{and}
  \bibinfo{author}{\bibfnamefont{A.}~\bibnamefont{Melchiorri}},
  \bibinfo{journal}{Phys. Rept.} \textbf{\bibinfo{volume}{364}},
  \bibinfo{pages}{1} (\bibinfo{year}{2002}), \eprint{astro-ph/0110348}.

\bibitem[{\citenamefont{Hu and White}(1997)}]{Hu:1997hp}
\bibinfo{author}{\bibfnamefont{W.}~\bibnamefont{Hu}} \bibnamefont{and}
  \bibinfo{author}{\bibfnamefont{M.~J.} \bibnamefont{White}},
  \bibinfo{journal}{Phys. Rev.} \textbf{\bibinfo{volume}{D56}},
  \bibinfo{pages}{596} (\bibinfo{year}{1997}), \eprint{astro-ph/9702170}.

\bibitem[{\citenamefont{Bogomol'nyi}(1976)}]{Bogomolnyi:1976}
\bibinfo{author}{\bibfnamefont{E.}~\bibnamefont{Bogomol'nyi}},
  \bibinfo{journal}{Sov. J. Nucl. Phys.} \textbf{\bibinfo{volume}{24}},
  \bibinfo{pages}{449} (\bibinfo{year}{1976}).

\bibitem[{\citenamefont{Kajantie et~al.}(1998)\citenamefont{Kajantie,
  Karjalainen, Laine, Peisa, and Rajantie}}]{Kajantie:1998bg}
\bibinfo{author}{\bibfnamefont{K.}~\bibnamefont{Kajantie}},
  \bibinfo{author}{\bibfnamefont{M.}~\bibnamefont{Karjalainen}},
  \bibinfo{author}{\bibfnamefont{M.}~\bibnamefont{Laine}},
  \bibinfo{author}{\bibfnamefont{J.}~\bibnamefont{Peisa}}, \bibnamefont{and}
  \bibinfo{author}{\bibfnamefont{A.}~\bibnamefont{Rajantie}},
  \bibinfo{journal}{Phys. Lett.} \textbf{\bibinfo{volume}{B428}},
  \bibinfo{pages}{334} (\bibinfo{year}{1998}), \eprint{hep-ph/9803367}.

\bibitem[{\citenamefont{Scherrer and Vilenkin}(1998)}]{Scherrer:1997sq}
\bibinfo{author}{\bibfnamefont{R.~J.} \bibnamefont{Scherrer}} \bibnamefont{and}
  \bibinfo{author}{\bibfnamefont{A.}~\bibnamefont{Vilenkin}},
  \bibinfo{journal}{Phys. Rev.} \textbf{\bibinfo{volume}{D58}},
  \bibinfo{pages}{103501} (\bibinfo{year}{1998}), \eprint{hep-ph/9709498}.

\bibitem[{\citenamefont{Rajantie}(2003)}]{Rajantie:2003xh}
\bibinfo{author}{\bibfnamefont{A.}~\bibnamefont{Rajantie}}
  (\bibinfo{year}{2003}), \eprint{hep-ph/0311262}.

\bibitem[{\citenamefont{Vincent et~al.}(1997)\citenamefont{Vincent, Hindmarsh,
  and Sakellariadou}}]{Vincent:1996qr}
\bibinfo{author}{\bibfnamefont{G.~R.} \bibnamefont{Vincent}},
  \bibinfo{author}{\bibfnamefont{M.}~\bibnamefont{Hindmarsh}},
  \bibnamefont{and}
  \bibinfo{author}{\bibfnamefont{M.}~\bibnamefont{Sakellariadou}},
  \bibinfo{journal}{Phys. Rev.} \textbf{\bibinfo{volume}{D55}},
  \bibinfo{pages}{573} (\bibinfo{year}{1997}), \eprint{astro-ph/9606137}.

\bibitem[{\citenamefont{Freedman et~al.}(2001)}]{Freedman:2000cf}
\bibinfo{author}{\bibfnamefont{W.~L.} \bibnamefont{Freedman}}
  \bibnamefont{et~al.}, \bibinfo{journal}{Astrophys. J.}
  \textbf{\bibinfo{volume}{553}}, \bibinfo{pages}{47} (\bibinfo{year}{2001}),
  \eprint{astro-ph/0012376}.

\bibitem[{\citenamefont{Kirkman et~al.}(2003)\citenamefont{Kirkman, Tytler,
  Suzuki, O'Meara, and Lubin}}]{Kirkman:2003uv}
\bibinfo{author}{\bibfnamefont{D.}~\bibnamefont{Kirkman}},
  \bibinfo{author}{\bibfnamefont{D.}~\bibnamefont{Tytler}},
  \bibinfo{author}{\bibfnamefont{N.}~\bibnamefont{Suzuki}},
  \bibinfo{author}{\bibfnamefont{J.~M.} \bibnamefont{O'Meara}},
  \bibnamefont{and} \bibinfo{author}{\bibfnamefont{D.}~\bibnamefont{Lubin}},
  \bibinfo{journal}{Astrophys. J. Suppl.} \textbf{\bibinfo{volume}{149}},
  \bibinfo{pages}{1} (\bibinfo{year}{2003}), \eprint{astro-ph/0302006}.

\bibitem[{\citenamefont{Knop et~al.}(2003)}]{Knop:2003iy}
\bibinfo{author}{\bibfnamefont{R.~A.} \bibnamefont{Knop}} \bibnamefont{et~al.}
  (\bibinfo{collaboration}{The Supernova Cosmology Project}),
  \bibinfo{journal}{Astrophys. J.} \textbf{\bibinfo{volume}{598}},
  \bibinfo{pages}{102} (\bibinfo{year}{2003}), \eprint{astro-ph/0309368}.

\bibitem[{\citenamefont{Landriau and Shellard}(2010)}]{Landriau:2010cb}
\bibinfo{author}{\bibfnamefont{M.}~\bibnamefont{Landriau}} \bibnamefont{and}
  \bibinfo{author}{\bibfnamefont{E.~P.~S.} \bibnamefont{Shellard}}
  (\bibinfo{year}{2010}), \eprint{1004.2885}.

\bibitem[{\citenamefont{Pogosian et~al.}(2010)\citenamefont{Pogosian, Tye,
  Wasserman, and Wyman}}]{Pogosian:2008amV2}
\bibinfo{author}{\bibfnamefont{L.}~\bibnamefont{Pogosian}},
  \bibinfo{author}{\bibfnamefont{S.~H.~H.} \bibnamefont{Tye}},
  \bibinfo{author}{\bibfnamefont{I.}~\bibnamefont{Wasserman}},
  \bibnamefont{and} \bibinfo{author}{\bibfnamefont{M.}~\bibnamefont{Wyman}}
  (\bibinfo{year}{2010}), \eprint{arXiv: 0804.0810v2 [astro-ph]}.

\bibitem[{\citenamefont{Lewis et~al.}(2000)\citenamefont{Lewis, Challinor, and
  Lasenby}}]{CAMB=Lewis:1999bs}
\bibinfo{author}{\bibfnamefont{A.}~\bibnamefont{Lewis}},
  \bibinfo{author}{\bibfnamefont{A.}~\bibnamefont{Challinor}},
  \bibnamefont{and} \bibinfo{author}{\bibfnamefont{A.}~\bibnamefont{Lasenby}},
  \bibinfo{journal}{Astrophys. J.} \textbf{\bibinfo{volume}{538}},
  \bibinfo{pages}{473} (\bibinfo{year}{2000}), \eprint{astro-ph/9911177}.

\bibitem[{\citenamefont{Sunyaev and Zeldovich}(1970)}]{Sunyaev:1970eu}
\bibinfo{author}{\bibfnamefont{R.~A.} \bibnamefont{Sunyaev}} \bibnamefont{and}
  \bibinfo{author}{\bibfnamefont{Y.~B.} \bibnamefont{Zeldovich}},
  \bibinfo{journal}{Astrophys. Space Sci.} \textbf{\bibinfo{volume}{7}},
  \bibinfo{pages}{3} (\bibinfo{year}{1970}).

\bibitem[{\citenamefont{Komatsu and Seljak}(2002)}]{Komatsu:2002wc}
\bibinfo{author}{\bibfnamefont{E.}~\bibnamefont{Komatsu}} \bibnamefont{and}
  \bibinfo{author}{\bibfnamefont{U.}~\bibnamefont{Seljak}},
  \bibinfo{journal}{Mon. Not. Roy. Astron. Soc.}
  \textbf{\bibinfo{volume}{336}}, \bibinfo{pages}{1256} (\bibinfo{year}{2002}),
  \eprint{astro-ph/0205468}.

\bibitem[{\citenamefont{Komatsu et~al.}(2010)}]{WMAP7-C:Komatsu:2010fb}
\bibinfo{author}{\bibfnamefont{E.}~\bibnamefont{Komatsu}} \bibnamefont{et~al.}
  (\bibinfo{year}{2010}), \eprint{1001.4538}.

\bibitem[{\citenamefont{Seljak and Slosar}(2006)}]{Seljak:2006hi}
\bibinfo{author}{\bibfnamefont{U.}~\bibnamefont{Seljak}} \bibnamefont{and}
  \bibinfo{author}{\bibfnamefont{A.}~\bibnamefont{Slosar}},
  \bibinfo{journal}{Phys. Rev.} \textbf{\bibinfo{volume}{D74}},
  \bibinfo{pages}{063523} (\bibinfo{year}{2006}), \eprint{astro-ph/0604143}.

\bibitem[{\citenamefont{Bevis et~al.}()\citenamefont{Bevis, Hindmarsh, Kunz,
  and Urrestilla}}]{Bevis:2010data}
\bibinfo{author}{\bibfnamefont{N.}~\bibnamefont{Bevis}},
  \bibinfo{author}{\bibfnamefont{M.}~\bibnamefont{Hindmarsh}},
  \bibinfo{author}{\bibfnamefont{M.}~\bibnamefont{Kunz}}, \bibnamefont{and}
  \bibinfo{author}{\bibfnamefont{J.}~\bibnamefont{Urrestilla}},
  \bibinfo{note}{in preparation}.

\bibitem[{\citenamefont{Avgoustidis and Shellard}(2006)}]{Avgoustidis:2005nv}
\bibinfo{author}{\bibfnamefont{A.}~\bibnamefont{Avgoustidis}} \bibnamefont{and}
  \bibinfo{author}{\bibfnamefont{E.~P.~S.} \bibnamefont{Shellard}},
  \bibinfo{journal}{Phys. Rev.} \textbf{\bibinfo{volume}{D73}},
  \bibinfo{pages}{041301} (\bibinfo{year}{2006}), \eprint{astro-ph/0512582}.

\bibitem[{\citenamefont{Hindmarsh
  et~al.}(2009{\natexlab{b}})\citenamefont{Hindmarsh, Ringeval, and
  Suyama}}]{Hindmarsh:2009qk}
\bibinfo{author}{\bibfnamefont{M.}~\bibnamefont{Hindmarsh}},
  \bibinfo{author}{\bibfnamefont{C.}~\bibnamefont{Ringeval}}, \bibnamefont{and}
  \bibinfo{author}{\bibfnamefont{T.}~\bibnamefont{Suyama}},
  \bibinfo{journal}{Phys. Rev.} \textbf{\bibinfo{volume}{D80}},
  \bibinfo{pages}{083501} (\bibinfo{year}{2009}{\natexlab{b}}),
  \eprint{0908.0432}.

\bibitem[{\citenamefont{Hindmarsh
  et~al.}(2009{\natexlab{c}})\citenamefont{Hindmarsh, Ringeval, and
  Suyama}}]{Hindmarsh:2009es}
\bibinfo{author}{\bibfnamefont{M.}~\bibnamefont{Hindmarsh}},
  \bibinfo{author}{\bibfnamefont{C.}~\bibnamefont{Ringeval}}, \bibnamefont{and}
  \bibinfo{author}{\bibfnamefont{T.}~\bibnamefont{Suyama}}
  (\bibinfo{year}{2009}{\natexlab{c}}), \eprint{0911.1241}.

\bibitem[{\citenamefont{Regan and Shellard}(2009)}]{Regan:2009hv}
\bibinfo{author}{\bibfnamefont{D.~M.} \bibnamefont{Regan}} \bibnamefont{and}
  \bibinfo{author}{\bibfnamefont{E.~P.~S.} \bibnamefont{Shellard}}
  (\bibinfo{year}{2009}), \eprint{0911.2491}.

\end{thebibliography}

\end{document}